\documentclass[useAMS,usenatbib,usegraphicx]{mn2e}

\usepackage[latin1]{inputenc}
\usepackage{amsmath}
\usepackage{amsfonts}
\usepackage{amssymb}
\usepackage[english]{babel}

\usepackage{lastpage}
\usepackage{times}

\newcommand{\etal}{et~al.\ }
\newcommand{\eg}{e.g.\ }

\newcommand{\kms}{km~s$^{-1}$}

\newcommand{\OI}{O~{\sc i}}

\newcommand{\NaI}{Na~{\sc i}}
\newcommand{\MgII}{Mg~{\sc ii}}

\newcommand{\SII}{S~{\sc ii}}

\newcommand{\SiII}{Si~{\sc ii}}

\newcommand{\CaII}{Ca~{\sc ii}}
\newcommand{\TiII}{Ti~{\sc ii}}

\newcommand{\FeII}{Fe~{\sc ii}}
\newcommand{\FeIII}{Fe~{\sc iii}}

\newcommand{\Fefs}{$^{56}$Fe}

\newcommand{\Nifs}{$^{56}$Ni}

\newcommand{\Dm}{$\Delta m_{15}(B)$}
\newcommand{\aap}{A\&A}
\newcommand{\mnras}{MNRAS}
\newcommand{\apj}{ApJ}
\newcommand{\apjs}{ApJS}
\newcommand{\apjl}{ApJ}
\newcommand{\aj}{AJ}
\newcommand{\pasp}{PASP}
\newcommand{\newar}{New Astron. Rev.}

\usepackage[colorlinks=false, pdfborder=0 0 0,dvipdfm]{hyperref}

\begin{document}

\title[EW ratios and blueshifts in SN Ia spectra]{Exploring the spectroscopic diversity of Type Ia Supernovae}

\author[Hachinger, Mazzali, Benetti]{S. Hachinger$^{1}$,
P. A. Mazzali$^{1,2}$,
S. Benetti$^{3}$\thanks{E-mail: stefano.benetti@oapd.inaf.it}\\
$^1$Max-Planck-Institut f\"ur Astrophysik, Karl-Schwarzschildstr.\ 1, 85748 Garching, Germany\\
$^2$Istituto Nazionale di Astrofisica-OATs, Via Tiepolo, 11, 34131 Trieste, Italy\\
$^3$Istituto Nazionale di Astrofisica-OAPd, vicolo dell'Osservatorio 5, 35122 Padova, Italy}

\date{Updated arXiv ver., 2006-06-23, corrections included. The definitive version is available at \href{http://www.blackwell-synergy.com}{www.blackwell-synergy.com}.}

\pubyear{2006}
\volume{}
\pagerange{}

\maketitle

\begin{abstract}

The velocities and equivalent widths (EWs) of a set of absorption features
are measured for a sample of 28 well-observed Type Ia supernovae (SNe Ia)
covering a wide range of properties. The values of these quantities at
maximum are obtained through interpolation/extrapolation and plotted against
the decline rate, and so are various line ratios. The SNe are divided
according to their velocity evolution into three classes defined
in a previous work of Benetti et al.: low velocity gradient (LVG),
high velocity gradient (HVG), and FAINT. It is found that all the LVG SNe have
approximately uniform velocites at B maximum, while the FAINT SNe
have values that decrease with increasing \Dm, and the HVG SNe have a
large spread. The EWs of the Fe-dominated features are approximately
constant in all SNe, while those of intermediate mass element (IME) lines
have larger values for intermediate decliners and smaller values for
brighter and FAINT SNe. The HVG SNe have stronger \SiII\ 6355-\AA\
lines, with no correlation with \Dm. It is also shown that the 
\SiII\ 5972\,\AA\ EW and three EW ratios, including one analogous 
to the $\cal R$(\SiII) ratio introduced by Nugent et al., 
are good spectroscopic indicators of luminosity. 
The data suggest that all LVG SNe have approximately constant kinetic 
energy, since burning to IME extends to similar velocities. The 
FAINT SNe may have somewhat lower energies. The large velocities and 
EWs of the IME lines of HVG SNe appear correlated with each other, 
but are not correlated with the presence of high-velocity features 
in the \CaII\ infrared triplet in the earliest spectra for the SNe for 
which such data exist.

\end{abstract}


\section{Introduction}

Understanding the physics of Type Ia Supernova (SN Ia) explosions is one of the
most important issues of contemporary astrophysics, given the role SNe Ia play
as distance indicators for cosmology and as main producers of heavy elements in
the Universe. One of the keys to enter the secrets of SNIa explosions is
to study their diversity.

SNe Ia are thought to be the thermonuclear explosion of carbon--oxygen white
dwarfs driven to ignition conditions by accretion in a binary system. Since
explosive burning of the CO mixture occurs when the white dwarf's mass is close
to the Chandrasekhar limit, it has long been speculated that SNe~Ia should be
good candidates for standard candles. 

While earlier impressions were that SNe Ia are quite homogeneous, it was later
noted that there are intrinsic differences among them. However,
correlations have been found that describe SNe Ia as a one-parameter family.
\citet{phi93} measured the decline in $B$-band magnitude from $B$-band maximum
to 15 d later, a quantity he called $\Delta m_{15}(B)$, and found that
brighter objects have a smaller decline rate than dimmer ones. The decline rate
therefore not only is useful for arranging SNe Ia in a `photometric'
one-parameter sequence, but should also reflect, although possibly in a gross
way, SN Ia physics.

This is matched by a spectroscopic sequence \citep{nug95}, defined by $\cal
R$(\SiII), the ratio of the depth of two absorptions at 5800 and 6100\,\AA,
both of which are usually attributed to \SiII\ lines. This ratio correlates
with the absolute magnitude of SNe~Ia and, in turn, with the rate of decline.
Spectroscopic models suggest that most spectral differences are due to 
variations in the effective temperature. In the context of Chandrasekhar-mass
explosions, these variations can be interpreted in terms of a variation in the
mass of $^{56}$Ni produced in the explosion. The relative behaviour of the two
\SiII\ lines is, however, counterintuitive, and still lacks a thorough
theoretical explanation. Garnavich et al. (2004) suggest that the bluer line is
affected by \TiII\ lines for objects with $\Delta m_{15}(B)\geq1.2$, but this
is not supported by detailed spectral synthesis studies of SN~1991bg and
SN~2002bo (\citealt{maz97}; \citealt{ste05}).

Although a one-parameter description of SNe Ia has proved to be very
useful, it does not completely account for the observational diversity
of SNe Ia \citep[\eg][]{ben04,ben05}.  In fact, earlier studies
\citep[\eg][]{pat96,hat00} suggest that the photospheric expansion
velocity, which can be taken as a proxy for the kinetic energy release
in the explosion, correlates with neither $\Delta m_{15}(B)$ nor
$\mathcal{R}$(\SiII) [see also \citet{wel94} for an early attempt
to correlate SNIa observables]. However, \citet{ben05} found a spread
in the time-averaged rate of decrease of the expansion velocity of the \SiII\ 6355-\AA\ 
absorption after maximum, which might suggest another means of
classifying SNe Ia. They used $\langle\dot{v}\rangle$ among other
parameters to perform a computer-based hierarchical cluster analysis
of a sample of 26 SNe Ia. This led to a partitioning of the SNe into
three groups, called, respectively, high velocity gradient (HVG;
$\langle\dot{v}\rangle = 97\pm16$ km$\;$s$^{-1}\,$d$^{-1}$), low velocity gradient (LVG;
$\langle\dot{v}\rangle = 37\pm18$ km$\;$s$^{-1}\,$d$^{-1}$), and FAINT. The FAINT group
includes SNe that are intrinsically dim, on average $\sim 2$ mag
fainter than SNe belonging to the other two groups. Their velocity
gradient is large, $\langle\dot{v}\rangle = 87\pm20$ km$\;$s$^{-1}\,$d$^{-1}$. HVG and
LVG SNe have similar mean absolute blue magnitude at maximum, but the
HVG SNe have a smaller spread in $\Delta m_{15}(B)$, and all SNe~Ia with
$\Delta m_{15}(B) \le 1.05$ are LVGs. \citet{ben05} confirmed the
relation between $\cal R$(\SiII) and $\Delta m_{15}(B)$, but find a
larger scatter among LVG SNe, especially at the bright end.

Spectra are an invaluable source of information, on both kinematics
and the chemical composition of the ejecta. While other studies aim at
extracting information by reproducing spectra or line ratios with
models (Stehle et al. 2005; Bongard et al. 2005), this work is
based on direct measurements of spectroscopic data, providing
therefore a complementary approach [see also \citet{fol04}, who 
emphasized the time evolution]. We focus on a comparison of the spectral
properties of different SNe. This requires a large enough number 
of objects with spectral data of good quality and at comparable 
epochs of evolution. During the last few years, the number of such 
objects has increased tremendously.

This work is thus based upon a collection of published as well 
as unpublished spectral data for 28 SNe
Ia, 25 of which are from \citet{ben05}. The expansion velocities of some
clearly defined featues were measured systematically and consistently, as were
their equivalent widths (EWs, see section \ref{sec:EWMeasurementtechnique}). With the 
aim of gathering information about the
differences in chemical composition, we studied a number of line strength
ratios, involving lines of intermediate mass elements (IME) such as S or Si as
well as lines of Fe group elements. Thus, our study should explore the extent
of nuclear burning in different objects. It also turned out
to be interesting to look at ratios involving a line of oxygen as
representative of unburned or partially burned material. Taking into account seven 
different lines, this paper aims to extend the studies cited above.

Measurements were performed for different objects at comparable epochs of
evolution, so that values can be contrasted, differences examined and links to 
the physical properties of the objects identified. One possible approach is to
examine the relation between line velocities and strenghts on the one hand and
the light curve decline rate on the other.

\section{Spectral data and measurements}
\label{sec:measurements}

We used the sample of SNe from \citet{ben05}, except for one object (1999cw)
for which there was no spectum at a suitable epoch (see below). Three new
objects (1991M, 2004eo, 2005bl) were added; the sample was divided into the
three groups defined in \citet{ben05}. An overview of the objects, group
assignment, $\Delta m_{15}(B)$ and $\langle\dot{v}\rangle$ values, and the sources of the
respective spectra is given in Table \ref{tab:objtable}. These data are taken
from \citet{ben05} -- see also references therein -- except for additional/updated
objects (see table notes), for which they were newly calculated.

For our measurements, we selected from our database about 100 spectra containing
 at least one absorption that can be clearly identified.

\begin{figure*}
	\centering
	\includegraphics[width=0.58\textwidth,angle=270]{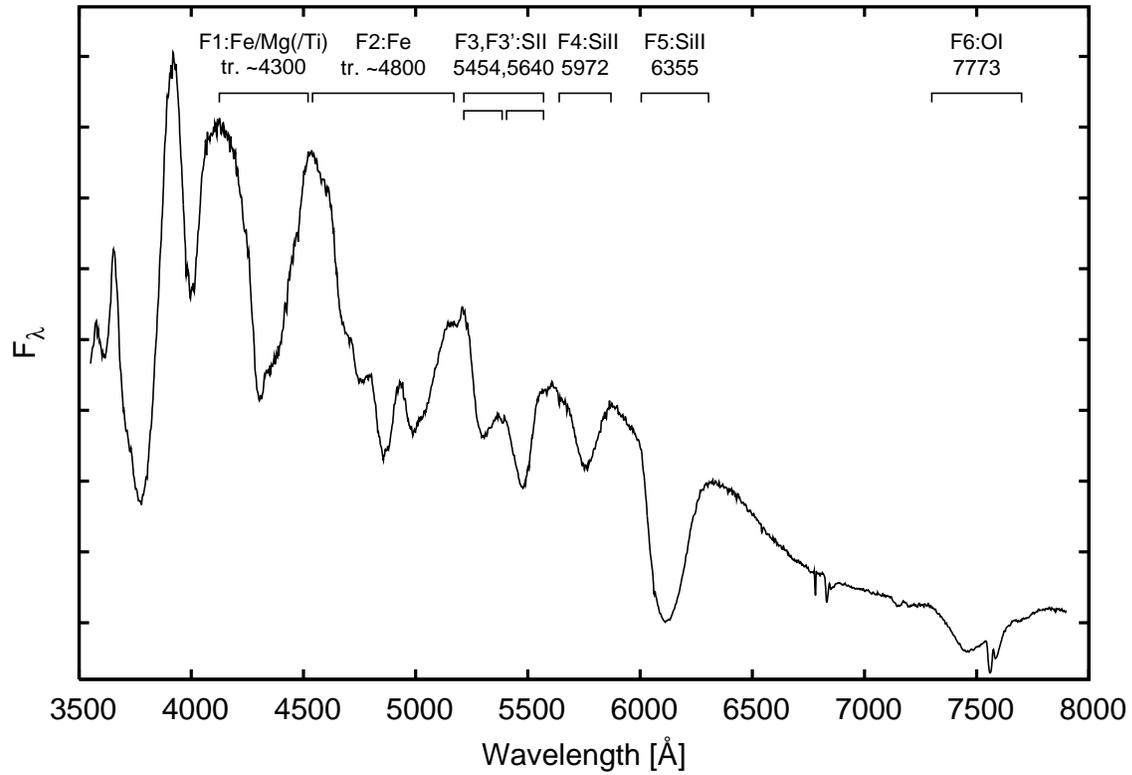}
	\caption{Overview of the features measured in an example spectrum.}
	\label{spectrum}
\end{figure*}

In a SN Ia spectrum near maximum, several such features can be seen; 
Table \ref{tab:featuretable} lists the lines used for this study. The EW ratios 
considered\footnote{Note that we use the EW as a measure for 
line strengths, while \citet{nug95} use line depth.} are given in Table 
\ref{tab:ratiotable}. For an overview of features, see Fig. \ref{spectrum}.

Starting from the blue end, a feature typical for SNe Ia shows up 
around $3750\,$\AA\, which is due to the \CaII\ H\&K lines. This feature 
is, however, not evaluated in this study because the spectra of some  
objects do not extend far enough to the blue to include that feature.

The first two features evaluated, viewed from the blue, are troughs (i.e.
unresolved blends of many lines) mostly related to Fe lines. The bluer one is
centered around $\sim4300\,$\AA\ in the observed spectra, and is made not only of
\FeII\ and \FeIII\  lines, but has also contributions from \MgII\, for the
fainter SNe and -- for very faint objects -- strong \TiII\ lines. The other 
trough is centered around $\sim4800\,$\AA\ in the spectra and contains lines  of
\FeII\ and \FeIII\ with almost no contamination by other elements except some
\SiII\ lines.  Both features are frequently accompanied by small `notches' at
their edges. These are weaker features that are not necessarily caused by Fe
lines.  The determination of the edges of the troughs is therefore difficult;
special care was taken that this was done in a consistent way for all
objects. Even then, measurements of the $\sim4300$-\AA\ trough for
the faintest objects cannot be compared with the rest; the appearance of \TiII\
lines affects not only the depth of the trough but also its blue extent. 
Also, measurements for 1991T-like objects should be taken with some
care because of the \FeIII\ lines dominating the feature before maximum
light (see \citealt*{maz95}).

Lines related to the IME ions \SII\ and \SiII\ are perhaps the most 
characteristic features of SN Ia spectra. \SII\ absorption causes a W-shaped
trough with observed minima at $\sim 5250$ and 5400\,\AA, respectively. The
minima can be attributed to \SII\ multiplets with their strongest lines at 
$\sim 5445$ and 5640\,\AA, respectively. All those transitions 
originate from relatively
high-lying lower levels, which causes a significant weakening of the \SII\
absorption at low temperatures. \SiII\ features, on the other hand, can be seen
at $\sim 5750$ and 6100\,\AA. Both features are blends, with average rest
wavelength 5972 and 6355\,\AA, respectively.  The redder line is by far the
stronger, and its temperature dependence is rather weak. On the other hand, the
strength of the 5972-\AA\ absorption is strongly correlated with temperature,
but the correlation is opposite to intuitive expectations from atomic physics
(see also section \ref{sec:EW_Si5972}).  An explanation might involve the
presence of other, weaker lines.

In the spectra of almost all SNe, another feature is
visible around $7500$\,\AA\ near maximum light, which is a blend of two very close
\OI\ lines with a mean rest wavelength of 7773\,\AA. The absorption is especially prominent
in spectra of FAINT SNe. Unfortunately, the feature suffers from contamination 
by the telluric absorption at $7605$\,\AA\ (and sometimes perhaps also by \MgII\ 
lines), which makes measurements complicated (see section \ref{sec:EWMeasurementtechnique}).

All measurements were carried out using {\sc iraf} (see Acknowledgments). The spectra 
were deredshifted; $z$ values were taken from wavelength measurements of interstellar 
\NaI\ D and \CaII\ H\&K lines, or -- if this was uncertain -- from either the literature 
if any was available or from the LEDA or NED catalogues (using galaxy recession
velocities). No reddening correction was applied to the spectra. As discussed
below, reddening should have negligible impact on the measured values.

The spectral data do not cover the same epochs for all objects. In order to
include as many SNe as possible, we interpolated/extrapolated EW and velocity
values at $t=0\;\textrm{d}$ for each object  by performing a least-squares fit of
the measured values at different times (this also yields standard deviations
used as statistical error estimates).  Since the features can be assumed to
evolve linearly in a limited interval of time, we evaluated spectra from --5 to
+5~d relative to $B$ maximum. In cases where only one or two spectra were
available in this range,  additional spectra from epochs between --8 and +6.5~d
were used if available  (for measurements of the \SiII\ $\lambda5972$ EW and
velocity, and of the  \SII\ EW, the upper limit was +5~d in all cases, as these
values evolve rapidly at later times for some objects).

After this selection, some objects remained with only one or two spectra
available; this required a special evaluation procedure. In cases with two
spectra, the error cannot be computed from the regression; it was therefore
calculated by propagating the errors of the single 
measurements\footnote{Quadratic error propagation using the derivatives of the
formula: $y_0=\left(y_1t_2-y_2t_1\right) / \left(t_2-t_1\right)$, where $y_0$ is the value at
$t=0$, and $y_{1,2};t_{1,2}$ are the coordinates of the two given data points.}
instead.  In cases with only one spectrum, the value from this spectrum is
given in the diagrams, if it can be expected that the evolution  between the
date of $B$ maximum and that of the spectrum is not too rapid. The error was then 
calculated by adding to the error of measurement (see below) an estimate for
the error in the estimated time. The latter was computed as the average slope
of the regression lines for objects belonging to the same SN group (with
$\geq3$ spectra available), multiplied by the time offset of the single
spectrum relative to the day of $B$ maximum.

For the SNe 1984A and 2002dj EW measurements, the described procedure is not 
reasonable, as the spectral coverage is sparse around $B$ maximum, and the observations
of these two objects show a particularly pronounced non-linear time evolution 
for many EWs. The EW values and respective uncertainties were thus obtained 
by performing a quadratic polynomial fit\footnote{Non-linear least-squares 
Marquardt--Levenberg algorithm; sign of the leading polynomial coefficient 
fixed manually after investigating data} to all available data between --10  
and +10~d relative to $B$ maximum.

Ratios were calculated from the $EW(t=0)$ values obtained  as described; the
errors attached to the ratios were computed by propagating the errors of the
EWs quadratically.

\subsection{Velocity measurements}

The mean expansion velocity of a given line was determined from the blueshift
of the absorption relative to the rest wavelength. The velocity thus derived is
physically meaningful only for single lines or for close multiplets that
are sufficiently isolated from other strong features. Since most lines have a 
P Cygni shape and are blends, one has to think thoroughly about how to
measure the `mean' wavelength of an absorption. We used two different methods: 
The first is to use the gaussian fit routine
within {\sc iraf}; the second is to estimate the centre of the absorption by eye
(taking into account problems such as the sloping continuum). The values
obtained from several such measurements were then averaged; the respective
standard deviation is a crude estimate of the error introduced by the manual
measurements, and was used to calculate the error in the cases mentioned above.
Fig. \ref{MessungIllustration} shows an example of how measurements were performed.

\subsection{Equivalent width measurements}
\label{sec:EWMeasurementtechnique}

As a measure of line strength, we take the equivalent width (EW). This is 
defined as:
\begin{eqnarray}
	EW=\int^{\lambda_1}_{\lambda_2}
	  \frac{F_{\rm C}(\lambda)-F(\lambda)}{F_{\rm C}(\lambda)} d\lambda, 
\end{eqnarray}
where $F(\lambda)$ is the flux density level in the spectrum and $F_{\rm C}(\lambda)$
the continuum flux density. EW is insensitive to multiplication 
of the flux density spectrum by a constant between $\lambda_1$ and $\lambda_2$. Therefore, 
if $\lambda_1$ and $\lambda_2$ are not too far away from each other, reddening 
effects can be neglected. Other `multiplicative errors' are also suppressed.

\begin{figure*}
	\centering
	\includegraphics[width=0.40\textwidth]{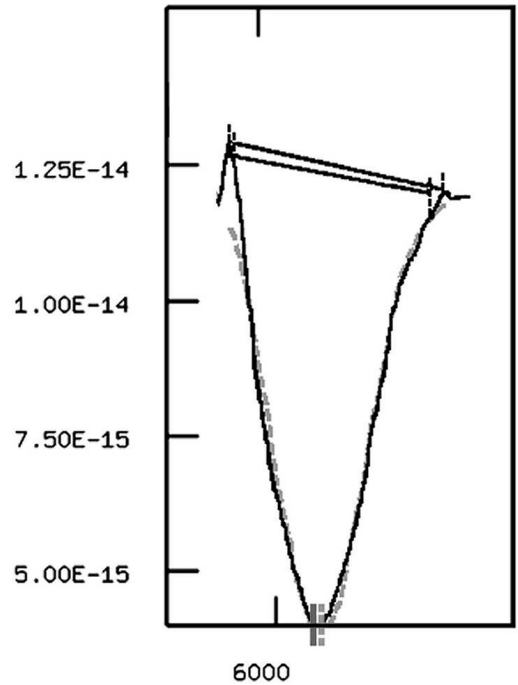}
	\caption{Examples for measurements: pseudo-continua are marked 
	in black; a gaussian fit and its centroid in light grey (dashed). A manual 
	centroid estimate is marked in dark grey.}
	\label{MessungIllustration}
\end{figure*}

The EW of an absorption (or an emission) line can be measured inside {\sc iraf} (when
showing spectra by \textsc{splot}) by entering the beginning and ending wavelengths of
the line, as well as the continuum level at those wavelengths. The main
difficulty is to define the continuum and the starting and ending points for a
feature, especially considering that the lines have P Cygni profiles. We
proceeded as follows (see Fig. \ref{MessungIllustration}):

Since a `real' continuum level cannot be determined in SN Ia spectra owing to
the multitude of line absorptions, for a single P Cygni profile we defined a
pseudo-continuum level to be the flux density level near the edges of the 
feature, neglecting the influence the emission component has on this. 
The edges of a feature were set roughly where the slope of the flux curve 
equals the slope of an imaginary pseudo-continuum curve joining the
opposite sides of the line. The error involved with this procedure has no
effect on the comparative study as long as all measurements are done
consistently. Some lines and absorption troughs regularly have poorly
determinable or jagged edges. In these cases, care was taken that the
measurements were done as homogeneously as possible for all objects.

EW measurements were carried out several times (see Fig.
\ref{MessungIllustration}), taking into account reasonable upper and lower
estimates of the continuum level and the starting and ending points. The
standard deviations thereby obtained for the EW values is again a rough estimate of
the errors introduced by the manual input of beginning/ending points and the
continuum level, and was further evaluated in the above-mentioned cases.

The \OI\ $\lambda7773$ line requires special attention. In most spectra, it is
contaminated by an atmospheric absorption that is not completely removed in the
reduction process. In these cases, the atmospheric absorption or its residuals
were cut by visual judgement before measuring the EW, and the measurements were
carried out eliminating the contamination in different ways, so that the
resulting standard deviation roughly represents the error introduced by the
snipping process.

\section{Expansion velocities}

We only measured the velocities of features that comply with the requirements
above, namely features that are due to a single ion, since single lines are not
available. The lines discussed below are \SiII\ $\lambda5972,6355$, and \SII\ 
$\lambda5640$.\footnote{The two \SII\ trough minima basically provide the same 
information, so we do not discuss measurements of the $\lambda5454$ 
minimum.} Measured values are listed in table \ref{tab:VelocityValueTable}.

\subsection{\SiII\ $\lambda6355$}

\begin{figure}
	\centering
	\includegraphics[width=0.58\textwidth, angle=270]{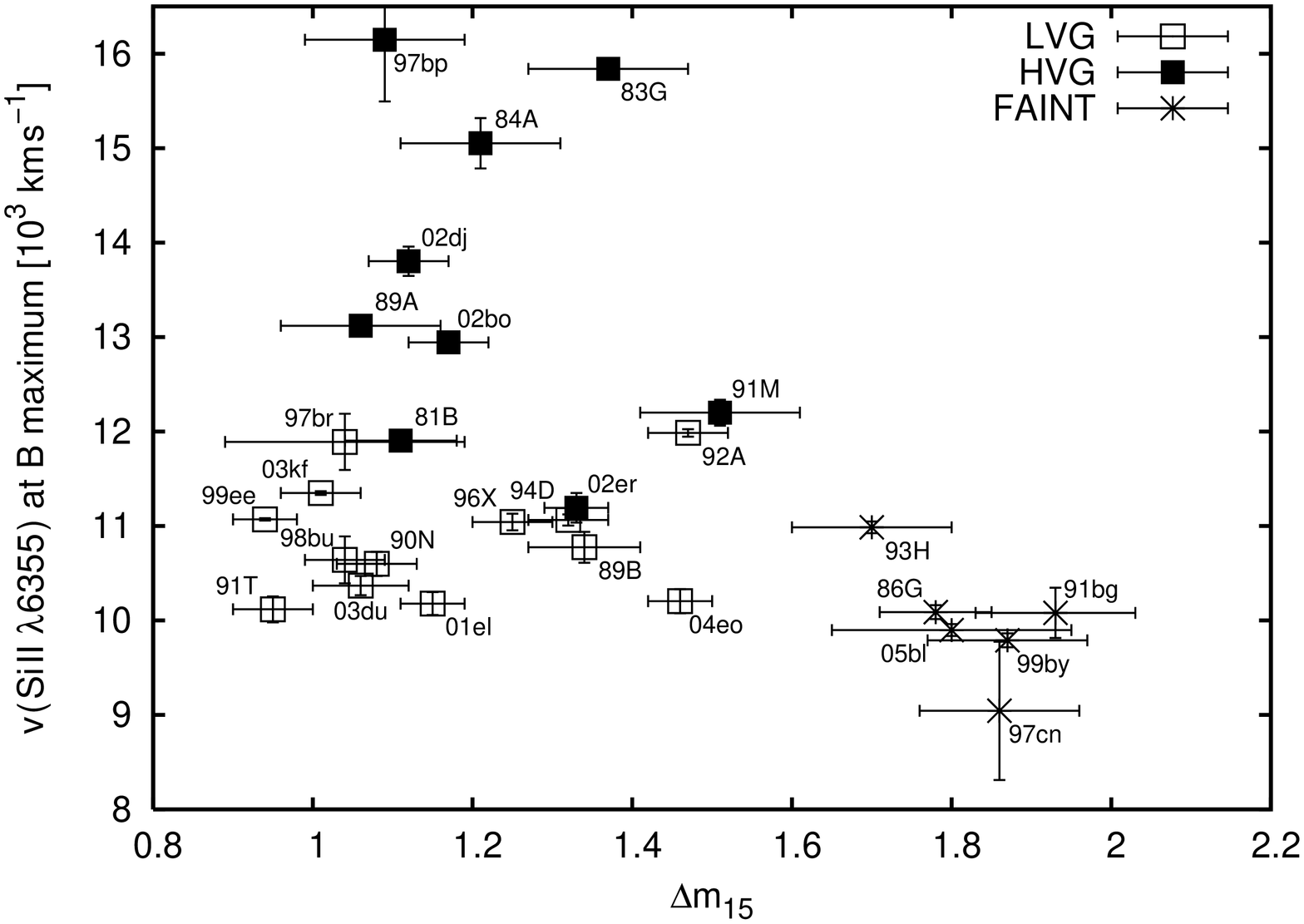}
	\caption{Expansion velocities from \SiII\ $6355$\,\AA\ blueshift 
		versus $\Delta m_{15}(B)$}
	\label{fig:V_Si6355}
\end{figure}

The velocities at maximum derived from this line show a big scatter 
especially at lower $\Delta m_{15}$ values (Fig. \ref{fig:V_Si6355}). 
However, once the SNe are divided  into velocity gradient groups, it 
turns out that most SNe, covering a wide range of $\Delta m_{15}$, 
from 0.9 to $\sim 1.7$, and including all LVG, some HVG, and the 
brightest among the FAINT SNe, have a roughly
constant $v$(\SiII\ $\lambda6355$), with only a small scatter ($11000 \pm
1000$\,\kms). HVG objects have a wide range of $v$(\SiII\ $\lambda6355$)
values, with no correlation with $\Delta m_{15}$, and are responsible for most
of the scatter. For FAINT SNe, $v$(\SiII\ $\lambda6355$) goes from values
comparable to those of the LVG group at $\Delta m_{15} \sim 1.5$--$1.7$ to
smaller values as $\Delta m_{15}$ increases.

\subsection{\SiII\ $\lambda5972$}

\begin{figure}
	\centering
	\includegraphics[width=0.58\textwidth, angle=270]{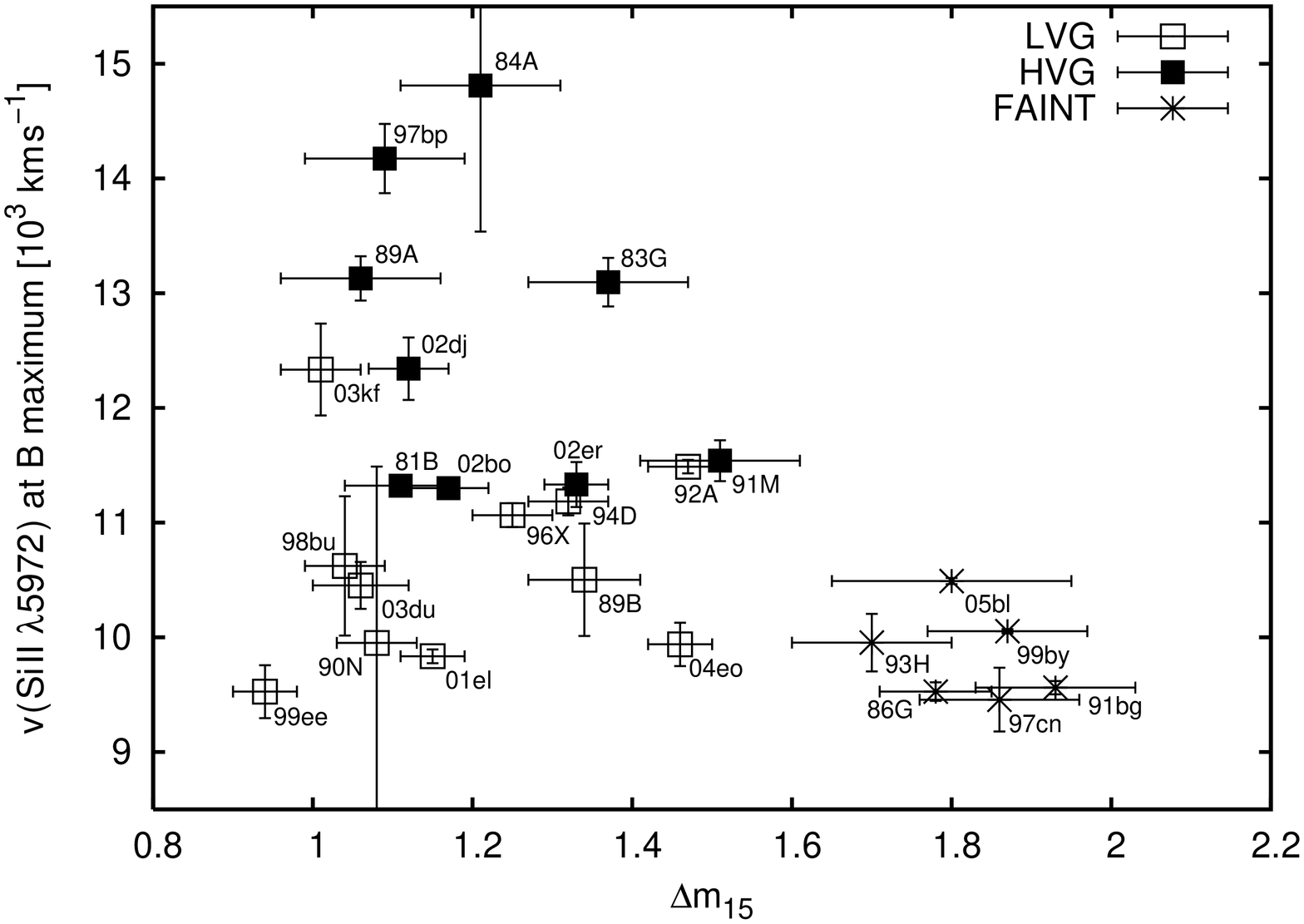}
	\caption{Expansion velocities from \SiII\ $5972$\,\AA\ blueshift 
		versus $\Delta m_{15}(B)$}
	\label{fig:V_Si5972}
\end{figure}

Although the velocities of this line (Fig. \ref{fig:V_Si5972}) show the same overall tendencies as those
derived from the $\lambda6355$ feature, there are differences
especially among the HVG objects. They reach lower maximum velocities, leading
to a smaller spread inside this group. A slight tendency to lower values (by
$\sim 500$\,\kms) can also be noted in every group. This is probably due to
the fact that the line is weaker and thus forms deeper than \SiII\
$\lambda6355$. The apparently larger scatter among LVG objects may be due to
the fact that the weak feature often shows a more complicated shape and suffers
from noise, making measurements of the centroid less reliable. Also,
contamination from other lines may occur.

\subsection{\SII\ $\lambda5640$}

\begin{figure}
	\centering
	\includegraphics[width=0.58\textwidth, angle=270]{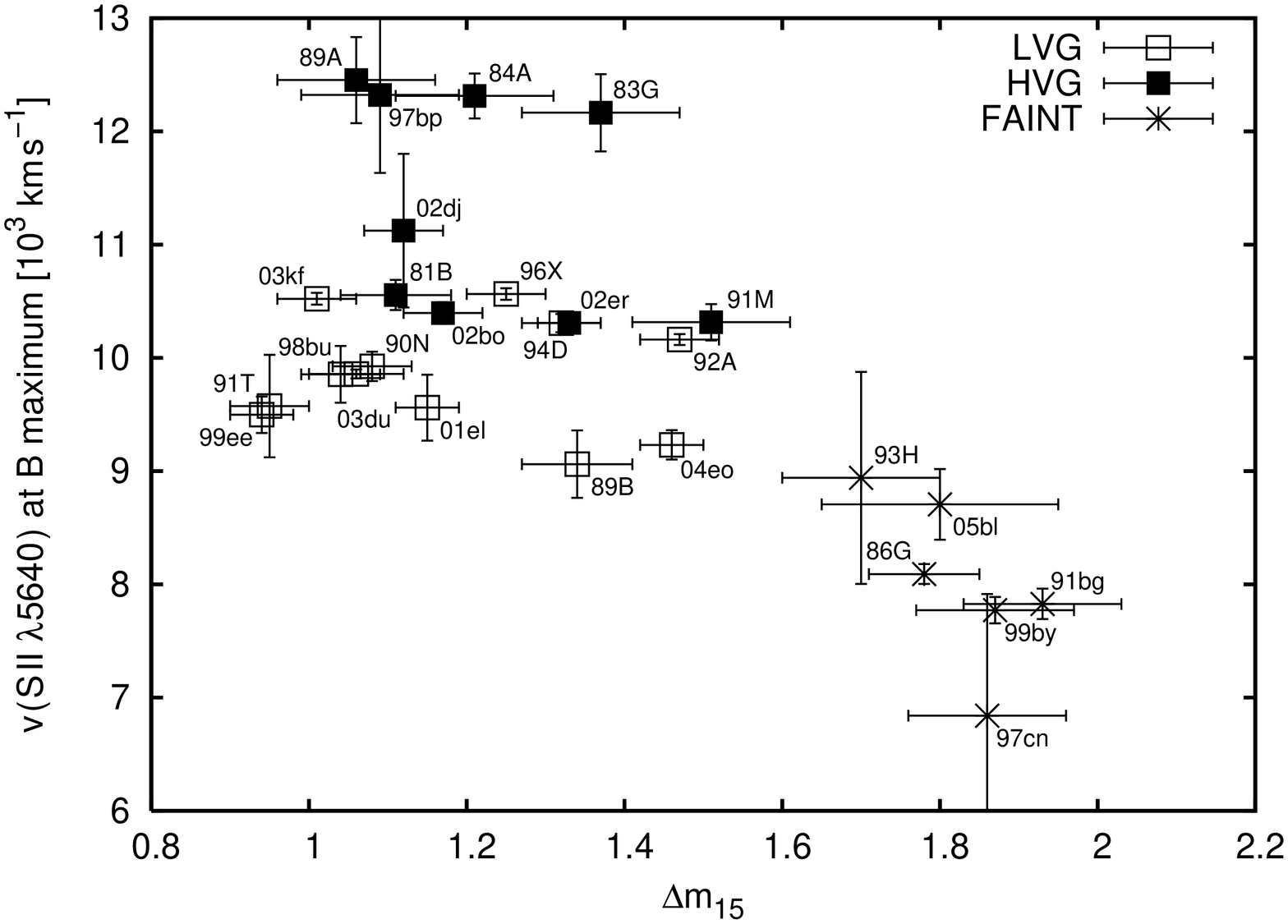}
	\caption{Expansion velocities from \SII\ $5640$\,\AA\ blueshift 
		versus $\Delta m_{15}(B)$}
	\label{fig:V_S5640}
\end{figure}

This \SII\ absorption (Fig. \ref{fig:V_S5640}) has a behaviour similar to
that of the \SiII\ absorptions discussed above. However, it shows significantly
lover velocities than the \SiII\ $\lambda6355$ feature, as can be expected
since the line is much weaker \citep[see also][]{blo06}. The mean differences
from the values derived from the \SiII\ line for the  respective groups are as
follows: HVG: $\sim 1000$--$4000$\,\kms; LVG: $\sim 1000$\,\kms; FAINT: $\sim
2000$\,\kms. The spread of values for the HVGs is much smaller in $v$(\SII) than
in $v$(\SiII\ $\lambda6355$), and also slightly smaller than that of
$v$(\SiII\ $\lambda5972$).

\section{Equivalent widths}

In this section, the measurements of the individual features are presented and 
discussed. Values are given in table \ref{tab:EWValueTable}.

\subsection{Fe-Mg(-Ti) trough $\sim 4300$\,\AA}

\begin{figure}
	\centering
	\includegraphics[width=0.58\textwidth, angle=270]{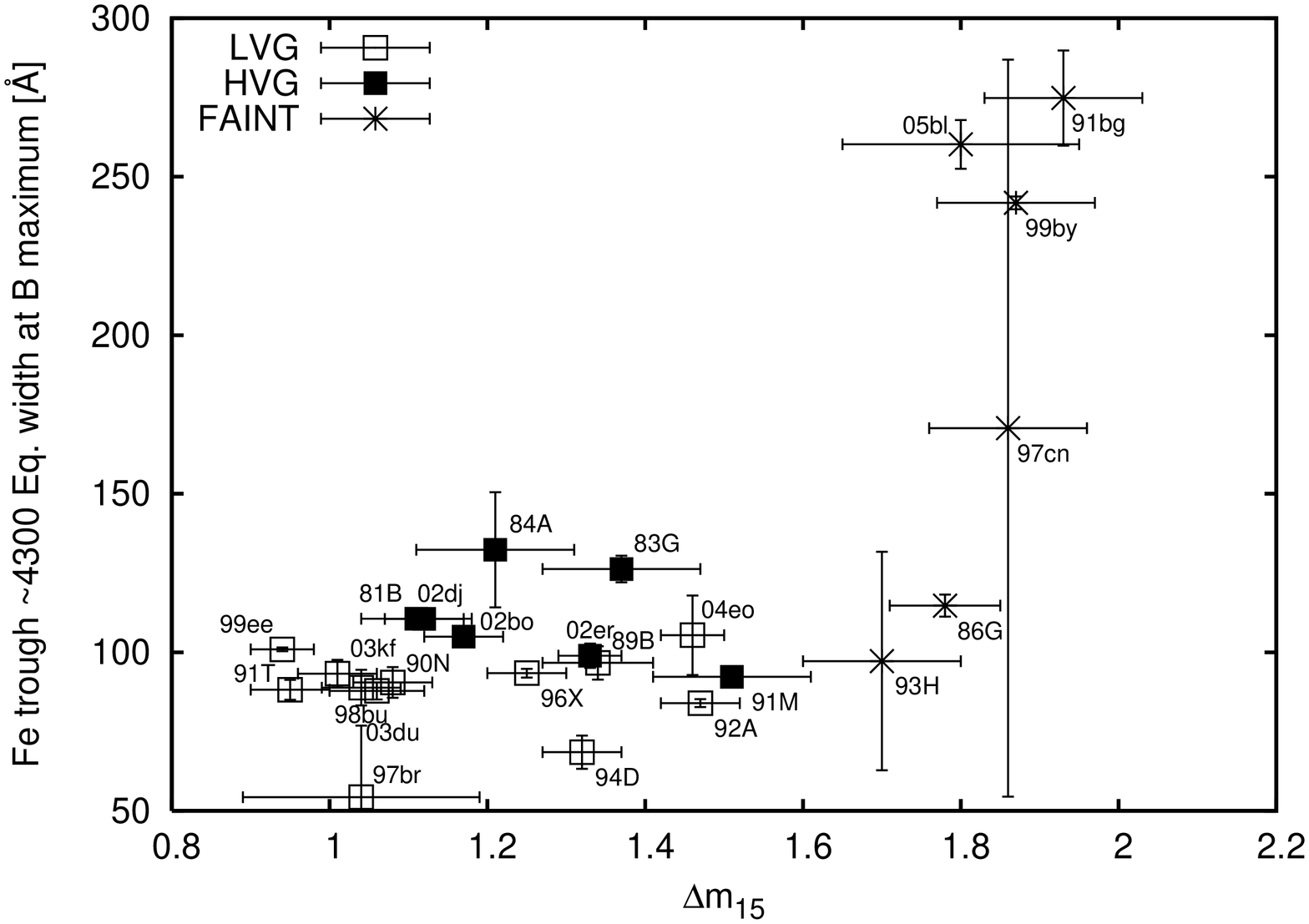}
	\caption{\FeII\ 4300\,\AA\ (observed wl.) EW versus $\Delta m_{15}(B)$}
	\label{fig:EW_Fe4300}
\end{figure}

The EW of this feature is roughly constant for all SNe with $\Delta m_{15}
\lesssim 1.8$ (Fig. \ref{fig:EW_Fe4300}). The lack of evolution suggests
that  Fe dominates this feature, or at least that the relative contribution of
Fe and Mg does not evolve with $\Delta m_{15}$ in the range from 1 to 1.8. HVGs
tend to have larger EWs than LVGs because in general they have broader and
deeper lines. The EW rapidly increases for FAINT SNe, which is the effect of
\TiII\ lines becoming very strong in those coolest objects, as was the case
for, \eg, SN~1991bg \citep{maz95}.

\subsection{Fe trough $\sim 4800$}

\begin{figure}
	\centering
	\includegraphics[width=0.58\textwidth, angle=270]{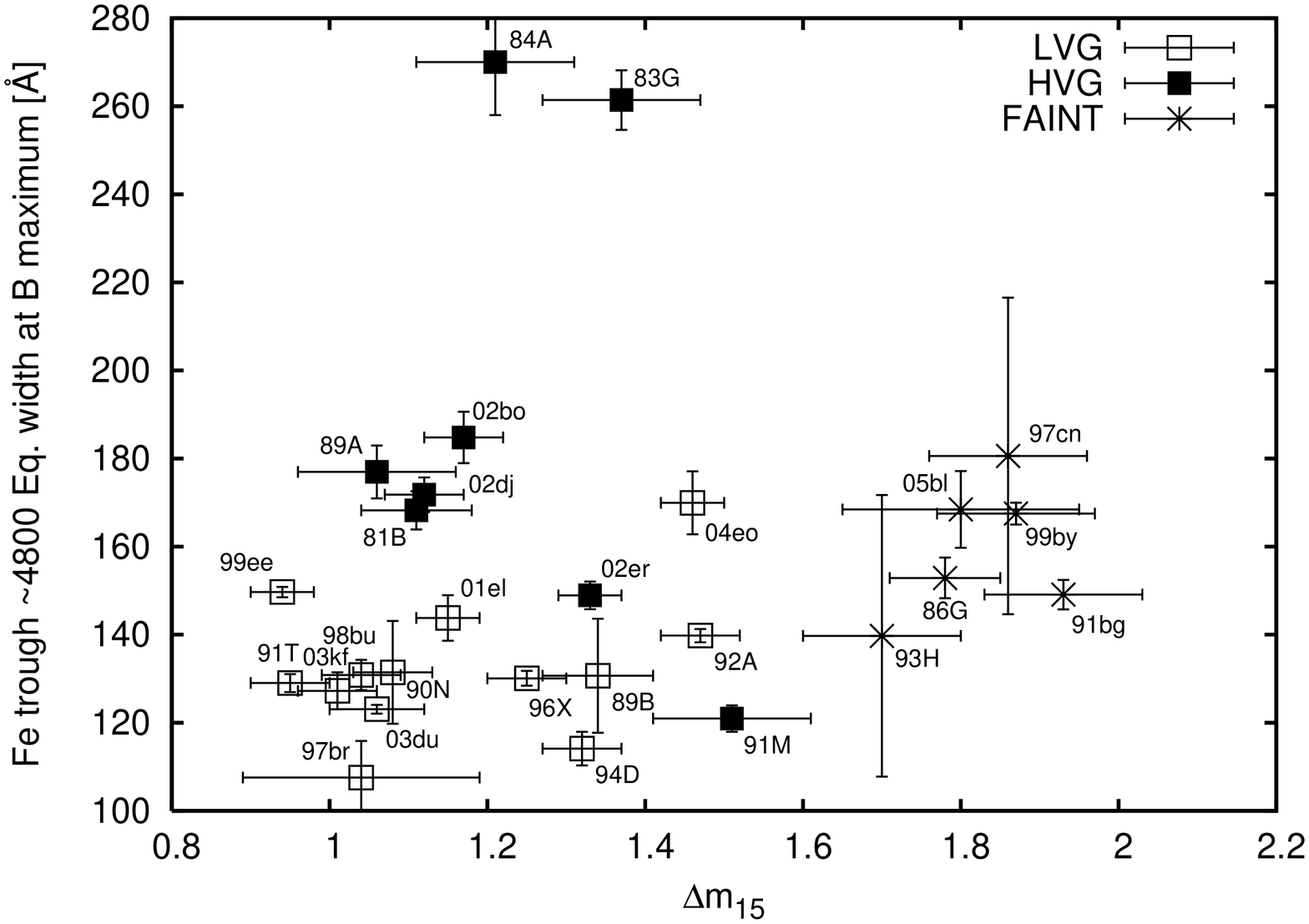}
	\caption{\FeII\ 4800\,\AA\ (observed wl.) EW versus $\Delta m_{15}(B)$}
	\label{fig:EW_Fe4800}
\end{figure}

The EW of this feature is essentially constant in all SNe, with a rather large
dispersion among objects with the same decline rate (Fig.
\ref{fig:EW_Fe4800}).  There is a slight trend to increasing values for the
fainter SNe, possibly an effect of the lower temperature which makes the \FeII\
lines stronger.  On the other hand, some of the peculiar bright SNe, such as
SN~1991T, where the \FeIII/\FeII\ ratio is large (Mazzali \etal 1995), have
values comparable to other SNe, suggesting that \FeIII\ dominates this feature,
as well as the Fe $\sim 4300$ trough, in all LVGs. HVG SNe have again larger values 
than LVGs, and now the trend is even
clearer. Analogy with the FAINT SNe may suggest that the HVG SNe have a lower temperature as a
consequence of the higher velocity, but it may also just imply that Fe reaches
higher velocities in HVGs, as do S and Si.

SNe~1984A and 1983G have somewhat larger values than the other SNe. This is
probably due to the broad-lined nature of these SNe, the SNe~Ia with the
highest velocities ever recorded \citep{ben05}. The other high-velocity SN,
SN~1997bp, could not be measured since its spectra do not extend to the blue. 

There is an apparent tendency for SNe to cluster in several small groups. We
refrain from interpreting this as an indication of different modes of the
explosion, and defer this to a time when more data are available.

\subsection{\SII\ trough $\lambda\sim5454,\sim5640$}
\label{sec:EW_S}

\begin{figure}
	\centering
	\includegraphics[width=0.58\textwidth, angle=270]{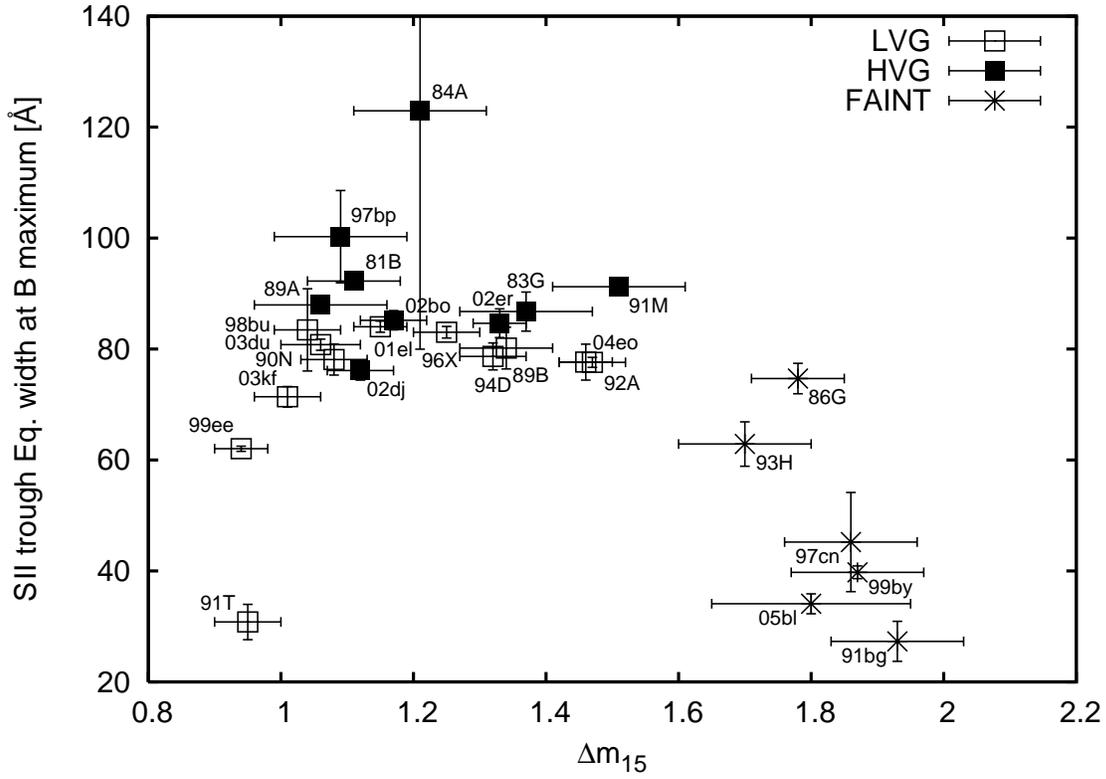}
	\caption{\SII\ trough EW versus $\Delta m_{15}(B)$}
	\label{fig:EW_S}
\end{figure}

The EW of the \SII\ feature (Fig. \ref{fig:EW_S}) shows a kind of parabolic 
trend, with very small scatter. It has a small value for SNe with $\Delta
m_{15} < 1.0$. It reaches a broad maximum in all other LVG and most HVG SNe with
\Dm$\;= 1.1$--$1.5$, and then it progressively declines at $\Delta m_{15} >1.6$. 
The observed drop may be explained as an effect of the insufficient population
of the highly excited lower levels of these lines  as the temperatures of the
SNe drop. However, at the highest temperatures a reduction in the IME abundance
is also required to reproduce the observed weakening of the lines in objects
such as SN~1991T (Mazzali \etal 1995). Therefore, it is possible that a trend
of increasing abundance going from the slowest to the intermediate decliners,
and then decreasing abundance from there to the fastest decliners is also
present.  The HVG SNe have a slightly larger value than the LVGs, and SN~1984A again 
stands out by having an anomalously large value.

\subsection{\SiII\ $\lambda5972$}
\label{sec:EW_Si5972}

\begin{figure}
	\centering
	\includegraphics[width=0.58\textwidth, angle=270]{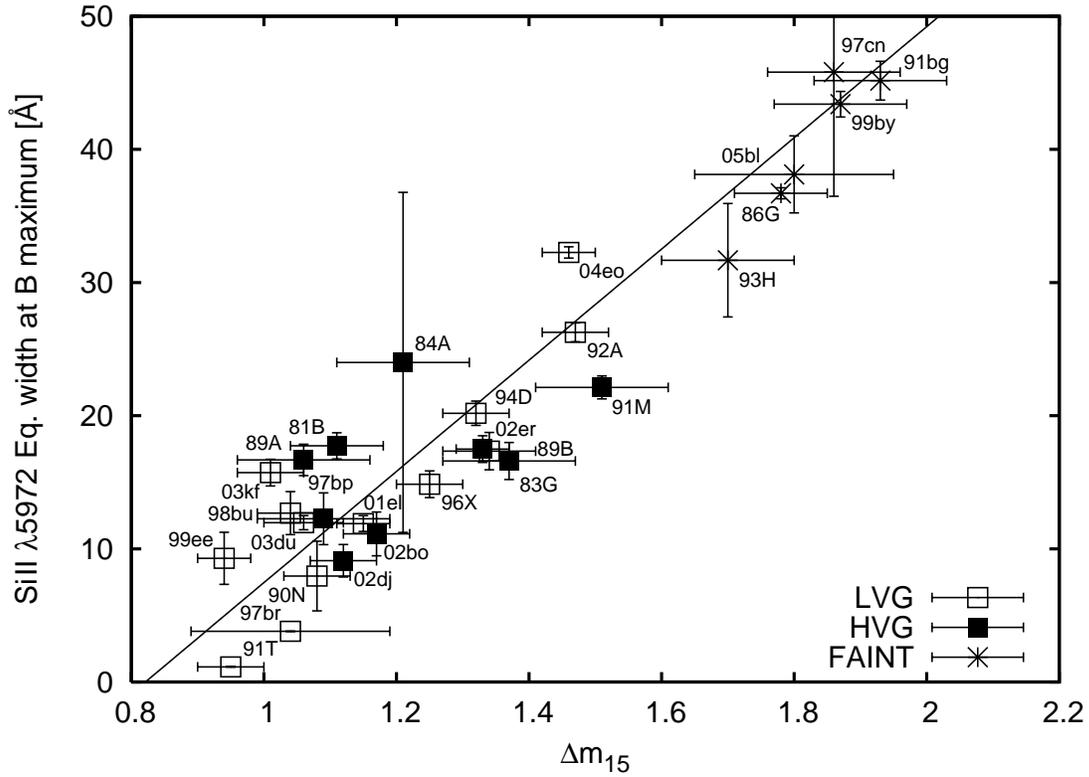}
	\caption{\SiII\ 5972\,\AA\ EW versus $\Delta m_{15}(B)$. Values for 91T and 97br are only 
	upper limit estimates (feature almost invisible).}
	\label{fig:EW_Si5972}
\end{figure}

\begin{figure}
	\centering
	\includegraphics[width=0.58\textwidth, angle=270]{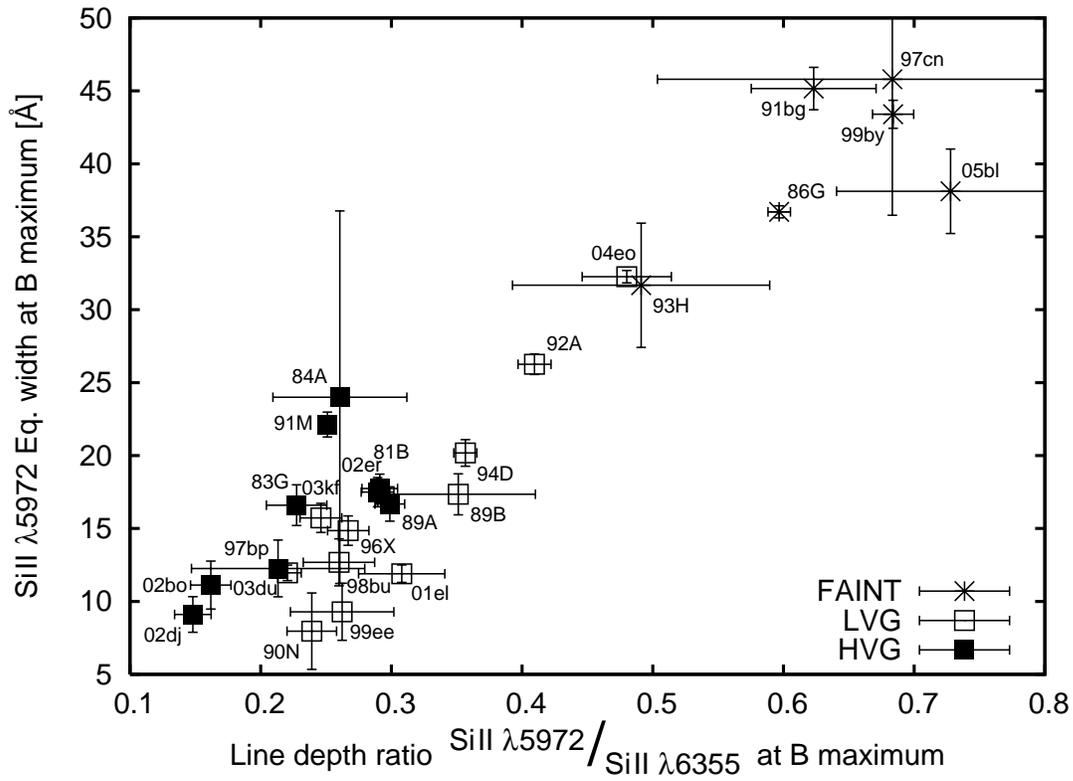}
	\caption{Comparison of our spectroscopic luminosity incdicator EW(\SiII\ 5972\,\AA) 
	to the ratio $\mathcal{R}$(\SiII), the quotient of the depths 
	of the \SiII\ 5972- and 6355-\AA\ features (\citep{nug95}). 
	Measurements and their analysis were carried out analogously to our EW measurements.}
	\label{fig:NugentComparison}
\end{figure}

\begin{figure}
	\centering
	\includegraphics[width=0.58\textwidth, angle=270]{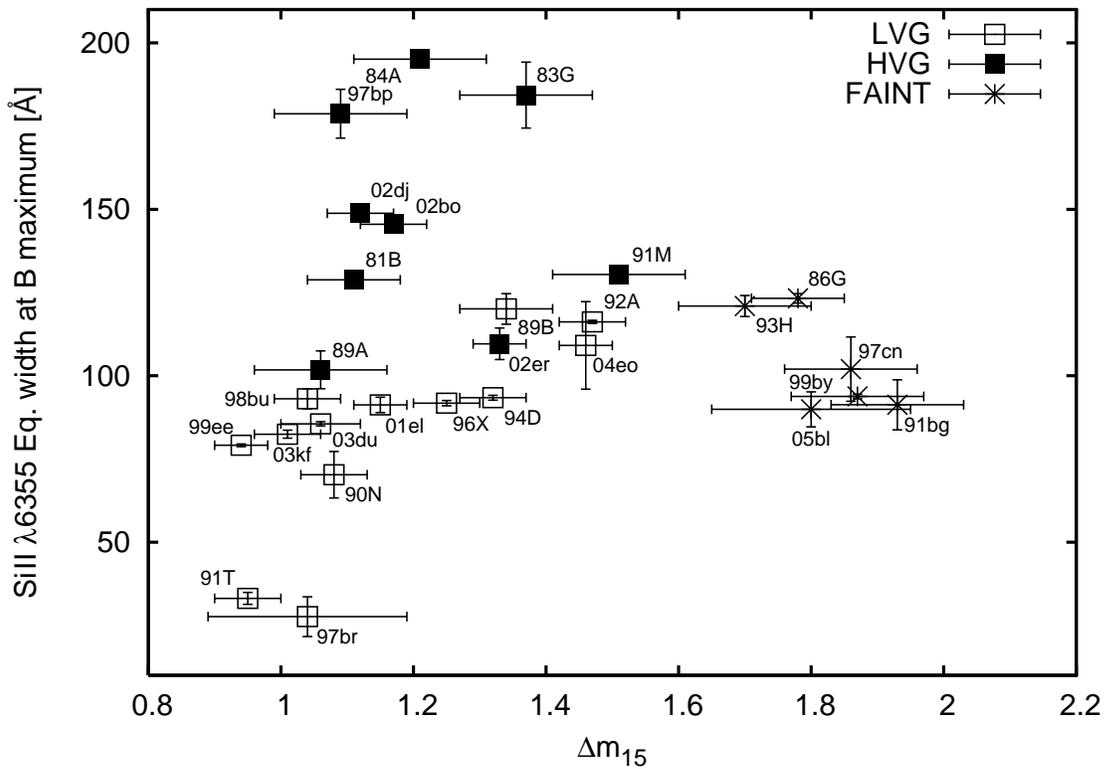}
	\caption{\SiII\ 6355\,\AA\ EW versus $\Delta m_{15}(B)$}
	\label{fig:EW_Si6355}
\end{figure}

The EW of the weaker \SiII\ line (Fig. \ref{fig:EW_Si5972}) correlates very 
well with $\Delta m_{15}$, and could therefore be used as a luminosity
indicator just as well as the line strength ratios presented below.  
This behaviour is at the basis of the observed relation between
$\mathcal{R}$(\SiII) and SN luminosity \citep{nug95}, as is illustrated 
by a plot of $\mathcal{R}$(\SiII) versus EW(\SiII\ $\lambda5972$) (Fig. 
\ref{fig:NugentComparison}) and by the weaker correlation of 
EW(\SiII\ $\lambda6355$) with $\Delta m_{15}$ (see Fig. 
\ref{fig:EW_Si6355}). The very existence of the trend 
is puzzling, since the \SiII\ $\lambda5972$ line originates from a
rather highly excited level, and its strength may be expected to correlate with
temperature directly rather than inversely. The explanation may involve the
contribution of lines from other elements and may require full non-local 
thermodynamic equilibrium (NLTE) analysis.
HVG SNe now blend in with the LVG SNe. This is somewhat surprising, since 
HVG SNe have the highest velocities (Fig. \ref{fig:V_Si6355}). Clearly, the 
line does not become more intense in SNe where it gets faster.

\subsection{\SiII\ $\lambda6355$}

This line shows a number of interesting trends (Fig. \ref{fig:EW_Si6355}). 
For most LVGs (with $\Delta m_{15} < 1.6$) the EW has a tendency to increase
slowly with increasing $\Delta m_{15}$. At larger decline rates, where the
FAINT SNe are, the value drops again. The two bright and peculiar LVG SNe,
1991T and 1997br, have much smaller values. This general behaviour is similar to
that of the \SII\ feature, and may be understood as the effect of temperature
and possibly of abundance: in SNe~1991T and 1999br the degree of ionisation is
higher than in spectroscopically normal objects, and the \SiII\ line is
accordingly weaker, but a low abundance of the IME is also required to
reproduce the observed spectra (Mazzali \etal 1995). The \SiII\ line is
strongest for intermediate decliners, where temperature reaches the optimal
value for this line and IME abundance possibly reaches a peak. The line weakens
in FAINT SNe, which are cooler and possibly have a smaller IME abundance.  This
effect is less marked than it is in the \SII\ feature, since the \SiII\
$\lambda6355$ originates from levels with a much smaller excitation potential
and is less sensitive to temperature. The observed drop may therefore more
directly reflect a change in the abundance of Si in near-photospheric mass
layers $(\sim 10000\,$\kms) of FAINT SNe.

The behaviour of the HVG SNe, on the other hand, is extremely different:  these SNe
are located almost vertically on the plot: although they cover a smaller
range of $\Delta m_{15}$ values than the LVG SNe (1.05 -- 1.5 versus 0.9 -- 1.5),
their EW(\SiII\ $\lambda6355$) spans about a factor of two in value. This may
reflect the presence of high-velocity absorption in the \SiII\ line
\citep{maz05a}. SN~1984A is again the most extreme object, followed by
SNe~1997bp and 1983G, but these objects appear to be the tip of a smooth
distribution.  The distribution of HVGs in EW is similar to that in $v$(\SiII\
$\lambda6355$) (Fig. \ref{fig:V_Si6355}). Faster lines tend to be broader and deeper.
Understanding this kind of behaviour may prove to be a very important step in
our effort to understand the systematics of SNe~Ia.

\subsection{\OI\ $\lambda7773$}

\begin{figure}
	\centering
	\includegraphics[width=0.58\textwidth, angle=270]{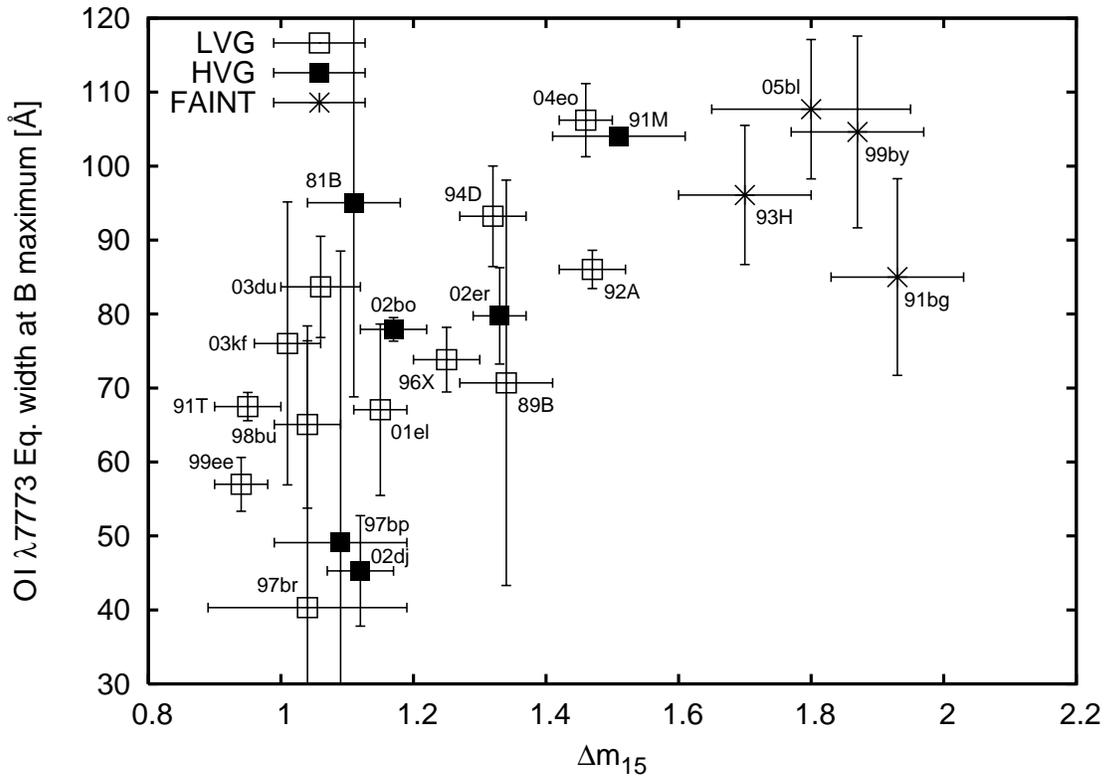}
	\caption{\OI\ 7773\,\AA\ EW versus $\Delta m_{15}(B)$. Note that for 90N this line 
	is too weak to be measured.}
	\label{fig:EW_O}
\end{figure}

The EW of this line (Fig. \ref{fig:EW_O}) tends to rise towards higher $\Delta
m_{15}$,  but shows quite a big scatter, especially at the bright end. Here,
there are both objects which show a very weak \OI\ feature around $B$ maximum
\footnote{90N is indeed missing in the plot because its \OI\ line is too weak to be 
measured; for other missing objects, no suitable spectral data in this wavelength 
range are available.} as well as objects
exhibiting values $\gtrsim90\,\textrm{\AA}$. Note that  these differences can be
found both within the HVG and LVG groups, which cover roughly  the same range of
measured \OI\ EW values. They may partly be due to the above-mentioned 
difficulties of measuring the \OI\ line. The overall trend of higher values 
for fainter objects is probably a temperature effect, but it may also reflect
changes in abundance. Among FAINT SNe, the trend appears to be reversed. This
is possibly due to the decrease of photospheric velocities at the faint end.

\section{Line strength ratios}

In this section we discuss selected ratios of EW. We focus on ratios that are
useful indicators of $\Delta m_{15}$, and on ratios that bear particular
physical significance because they involve elements that are synthesised in
different parts of the exploding white dwarf. The discussed ratio values
are given in table \ref{tab:RatioValueTable}.

\subsection{$\mathfrak{R}$(\SiII) (\SiII\ $\lambda5972$ versus \SiII\ $\lambda6355$)}

\begin{figure}
	\centering
	\includegraphics[width=0.58\textwidth, angle=270]{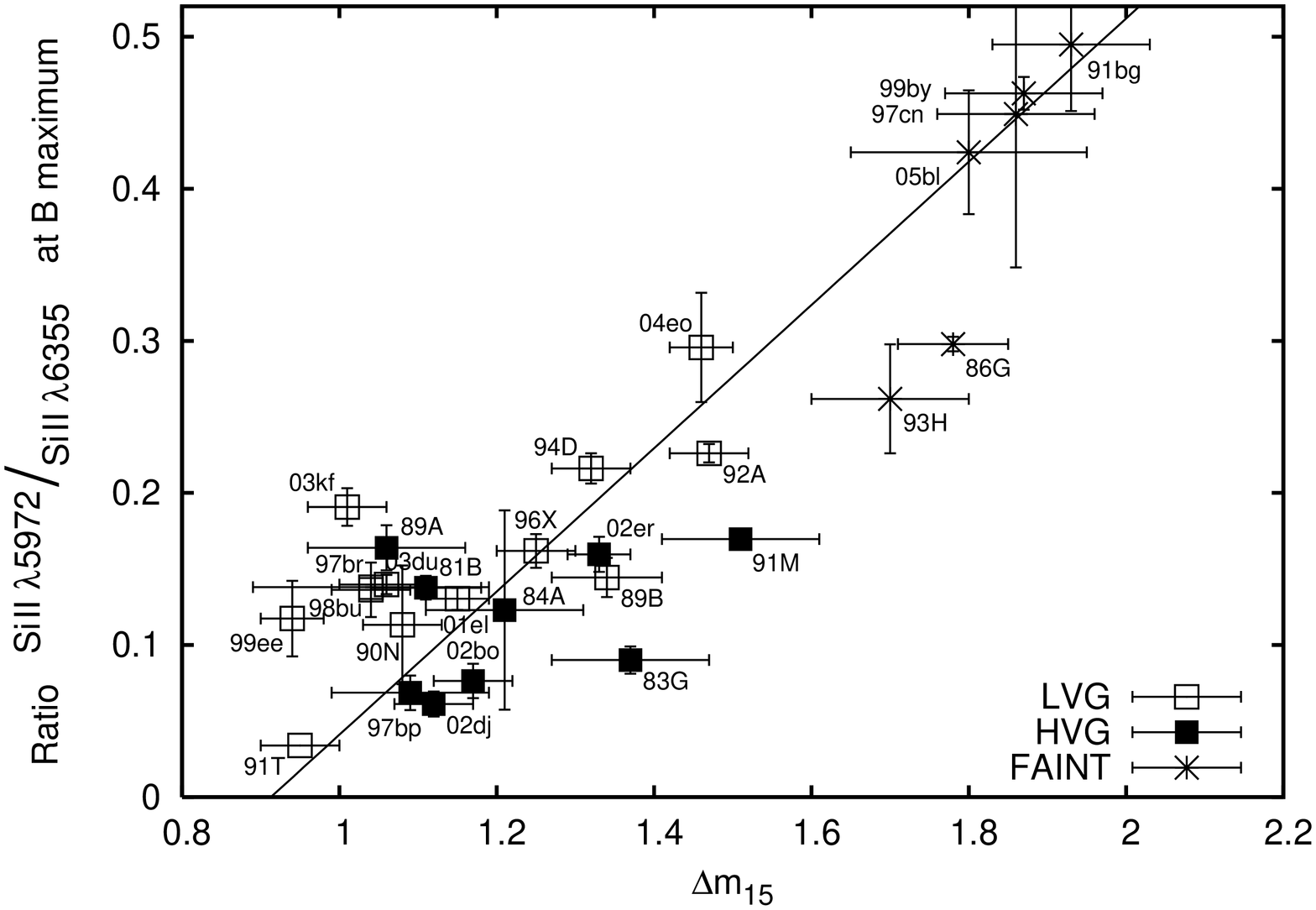}
	\caption{$\mathfrak{R}$(\SiII) -- EW(\SiII\ $5972$\,\AA)$\,/\,$EW(\SiII\ $6355\,$\AA) versus $\Delta m_{15}(B)$ (91T \& 97br: upper limits).}
	\label{fig:R_Si5972_Si6355}
\end{figure}

Our measurement is similar to the $\mathcal{R}$(\SiII) value of \citet{nug95},
but it differs from it since we use the EW. The EW ratio of the two \SiII\
lines follows the trend found by \citet{nug95} of increasing $\mathcal{R}$(Si
II) with increasing $\Delta m_{15}$ (Fig. \ref{fig:R_Si5972_Si6355}).
However, as noted in \citet{ben05}, the scatter at the bright end is larger. 
As we noted above, the observed behaviour is mainly caused by the
unexplained linear increase of the \SiII\ $\lambda5972$  line strength for
increasing $\Delta m_{15}$.

\subsection{Fe-Mg(-Ti) trough $\sim 4300$ versus Fe trough $\sim 4800$}

\begin{figure}
	\centering
	\includegraphics[width=0.58\textwidth, angle=270]{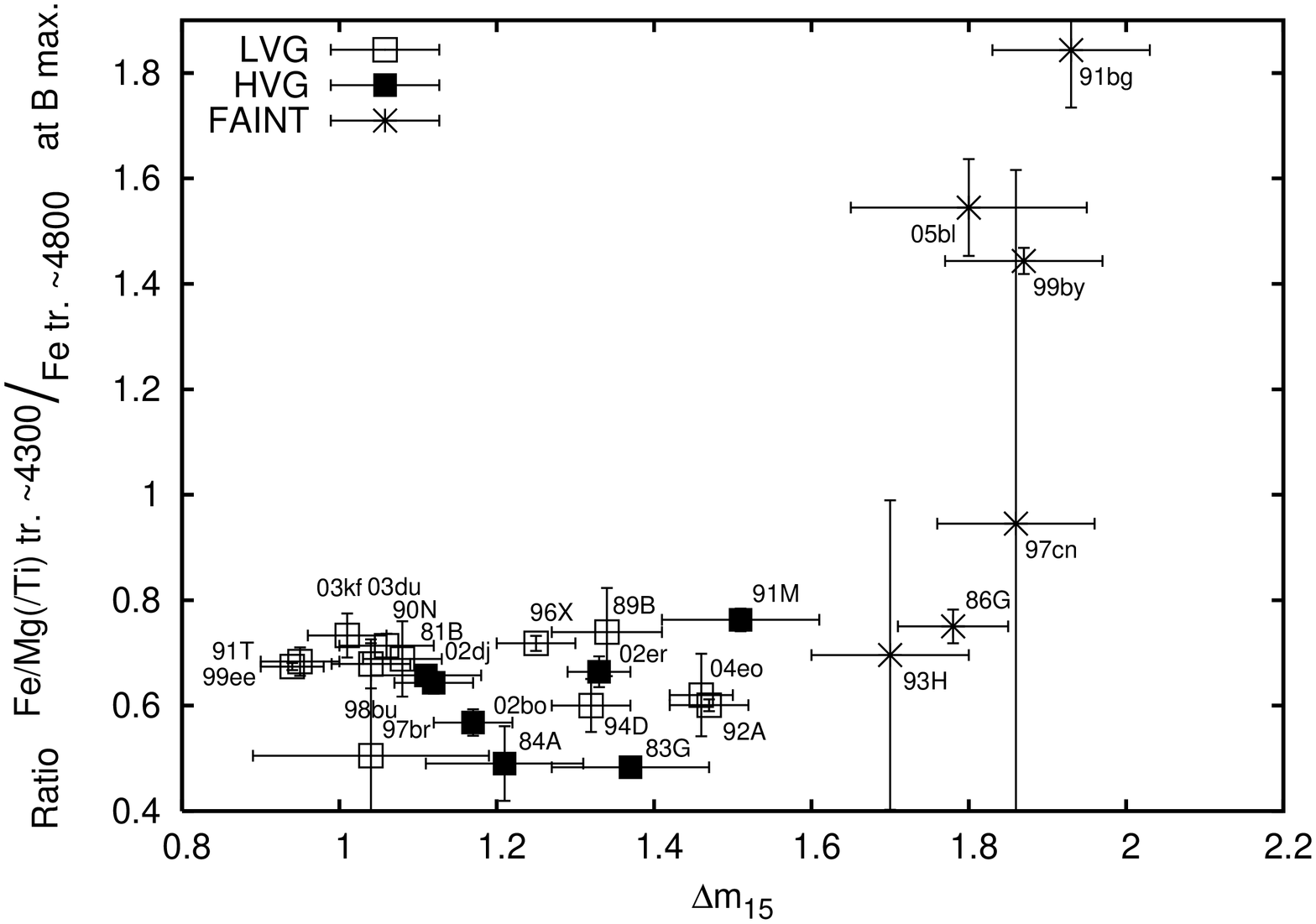}
	\caption{Ratio no. 5 -- EW(Fe trough 
  $\sim4300$\,\AA)$\,/\,$EW(Fe trough $\sim4800$\,\AA) versus $\Delta m_{15}(B)$: comparison of the Fe-dominated troughs.}
	\label{fig:R_Fe4300_Fe4800}
\end{figure}

The ratio of the EWs of these two broad absorption troughs (Fig. 
\ref{fig:R_Fe4300_Fe4800}) is fairly constant for $\Delta m_{15} \leq 1.8$. 
Some of the FAINT SNe (1991bg, 1999by, and 2005bl) have much larger values. 
The rise at the faint end is clearly due to the appearance of \TiII\ lines 
in the 4300-\AA\ feature at low temperature.

\subsection{\SII\ $\lambda\sim5454,\sim5640$ versus \SiII\ $\lambda6355$}

\begin{figure}
	\centering
	\includegraphics[width=0.58\textwidth, angle=270]{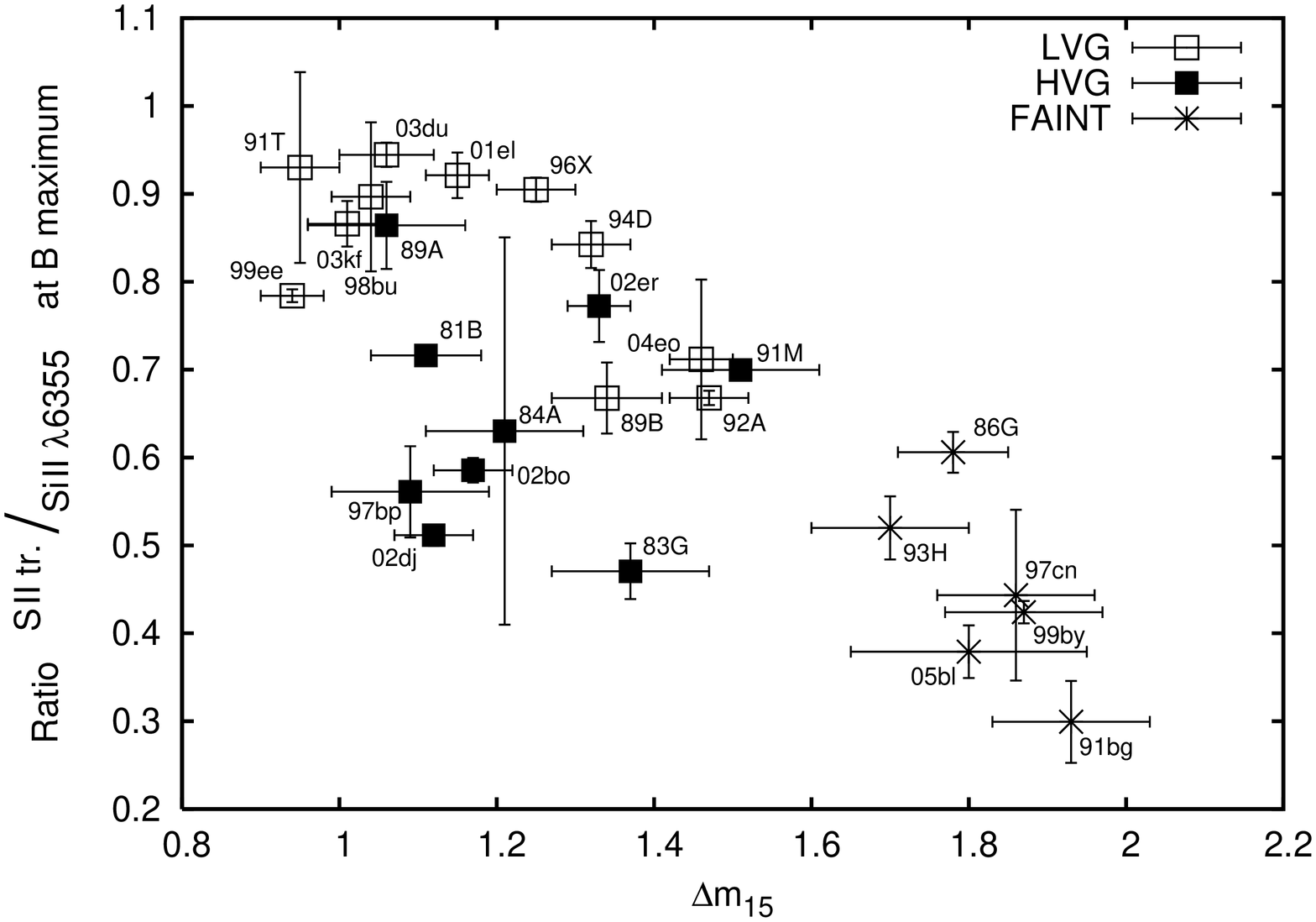}
	\caption{Ratio no. 4 -- EW(\SII\ trough)$\,/\,$EW(\SiII\ $6355$\,\AA) versus $\Delta m_{15}(B)$: comparison of IME.
		Note the HVG group behaviour.}
	\label{fig:R_S_Si6355}
\end{figure}

This value correlates very well with $\Delta m_{15}$ for FAINT objects (Fig.
\ref{fig:R_S_Si6355}). The LVG SNe also correlate reasonably well with $\Delta
m_{15}$, with a scatter of $\sim 10$ per cent, but the HVG SNe do not. The average values
of the HVG and the LVG group are very different. The HVG SNe show again an almost
vertical behaviour, as they did in both the $v$(\SiII\ $\lambda6355$) and
the EW(\SiII\ $\lambda6355$) plot. Since  EW(\SiII\ $\lambda6355$) is affected,
the ratio is smaller for these SNe. For fainter objects, the behaviour mainly
seems to reflect the above-mentioned (see section \ref{sec:EW_S}) changes of
ionization structure with decreasing temperature:  the \SII\ line strength
decreases rapidly as $\Delta m_{15}$ increases, which is not as much the case
for the Si line.

\subsection{$\mathfrak{R}$(S,Si) (\SII\ $\lambda\sim5454,\sim5640$ versus \SiII\ $\lambda5972$)}

\begin{figure}
	\centering
	\includegraphics[width=0.58\textwidth, angle=270]{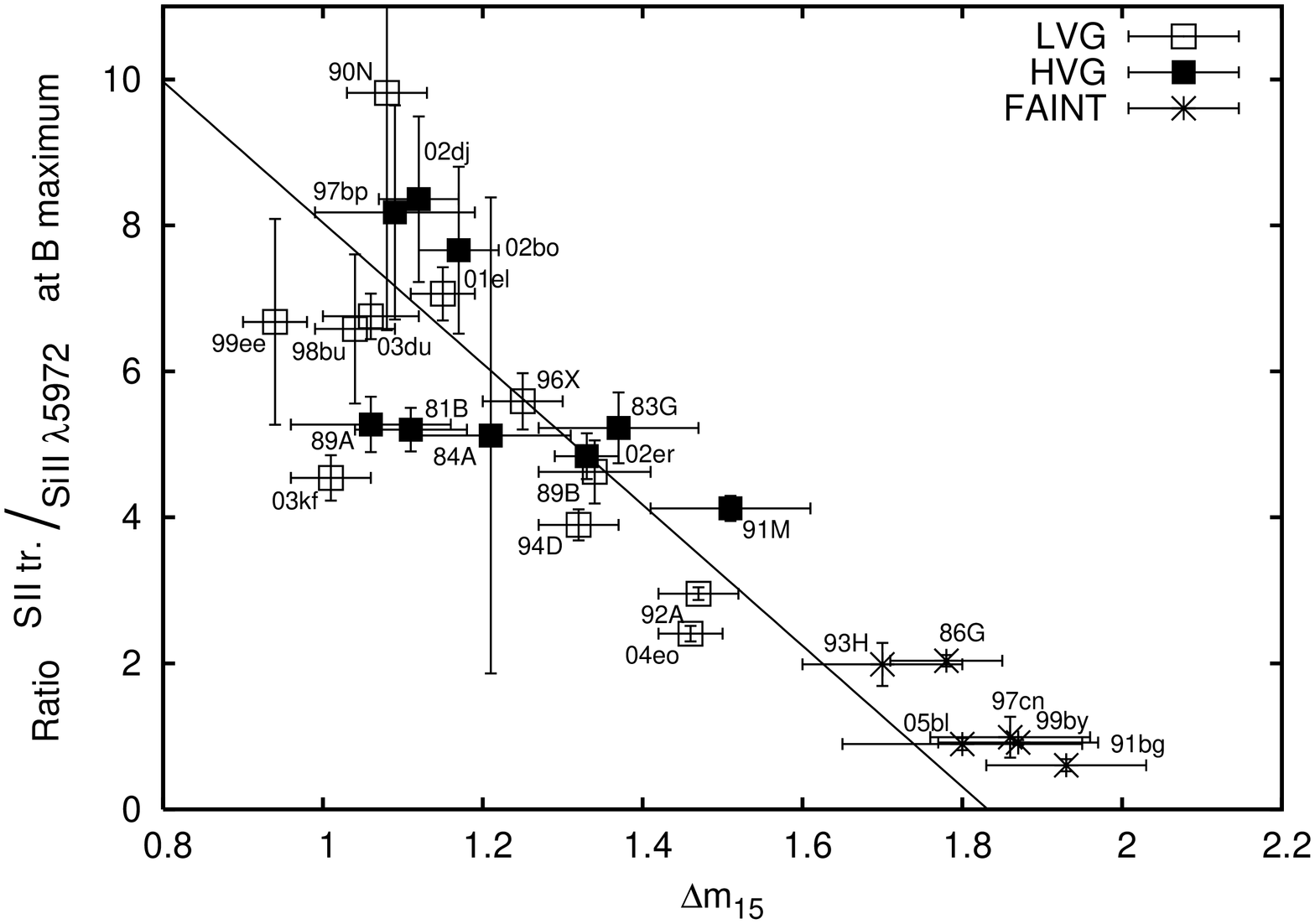}
	\caption{$\mathfrak{R}$(S,Si) -- EW(\SII\ trough)$\,/\,$EW(\SiII\ $5972$\,\AA) versus $\Delta m_{15}(B)$: `spectroscopic luminosity indicator'.}
	\label{fig:R_S_Si5972}
\end{figure}

This ratio correlates well with $\Delta m_{15}$ for almost all objects,
regardless of their group (Fig. \ref{fig:R_S_Si5972}). It decreases almost 
linearly with increasing $\Delta m_{15}$, and is thus as suitable as 
$\mathfrak{R}$(\SiII) as a spectroscopic luminosity indicator. The trend for a 
smaller ratio with increasing $\Delta m_{15}$ was already present in the
previous `S/Si' ratio, but here the scatter is much reduced and both LVG and
HVG objects follow the correlation, the differences between the two groups
being apparently suppressed.  These weaker lines are in fact less affected than
\SiII\ $\lambda6355$ by the high velocities and the ensuing increased strength,
as shown in the EW plots (Fig. \ref{fig:EW_S} and \ref{fig:EW_Si5972}). Even
SN~1984A follows the general trend: once ratios are taken its large EW values
cancel out. We cannot, however, draw any conclusions about Si distribution,
velocities, etc. from measurements involving the \SiII$\,\lambda5972$ feature,
because the behaviour of this line is not well understood, as discussed above.

\subsection{\SiII\ $\lambda6355$ versus Fe trough $\sim4800$}

\begin{figure}
	\centering
	\includegraphics[width=0.58\textwidth, angle=270]{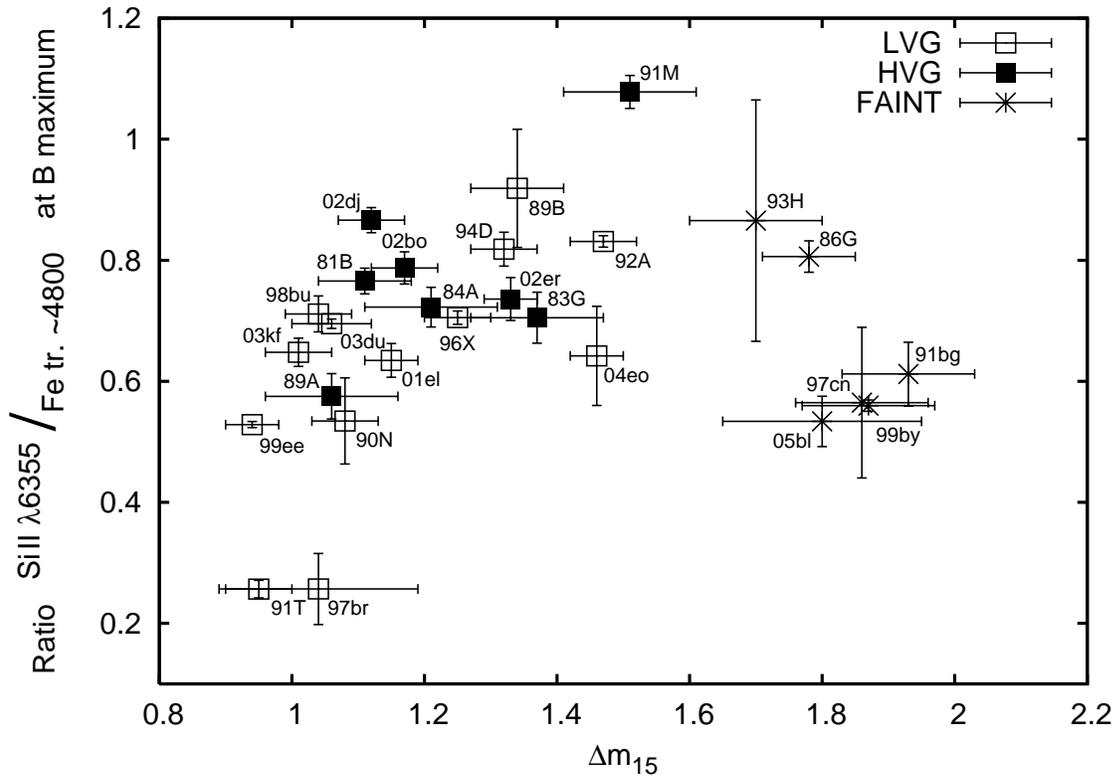}
	\caption{Ratio no. 2 -- EW(\SiII\ $6355$\,\AA)$\,/\,$EW(Fe trough $\sim4800$\,\AA) versus $\Delta m_{15}(B)$ -- IME feature versus Fe-dominated trough I. Note the drop at the faint end.}
	\label{fig:R_Si6355_Fe4800}
\end{figure}

The plot of this ratio (see Fig. \ref{fig:R_Si6355_Fe4800}) is 
very interesting, as is its possible meaning, which is discussed 
below. The ratio exhibits a `quadratic' behaviour: The values are 
small at small $\Delta m_{15}$, they increase until they reach a 
peak at $\Delta m_{15} \sim 1.1$--$1.5$ and then they drop again 
for very faint SNe such as 1991bg, 1997cn and 1999by. The
behaviour reflects that of EW(\SiII\ $\lambda6355$) but is highly enhanced,
suggesting that we are seeing more than just the effect of temperature. The HVG SNe
blend in with the other SNe, although they have larger values of both EW(\SiII\
$\lambda6355$) and EW(Fe $\sim4800$).

\subsection{\SII\ $\lambda\sim5454,\sim5640$ versus Fe trough $\sim4800$}

\begin{figure}
	\centering
	\includegraphics[width=0.58\textwidth, angle=270]{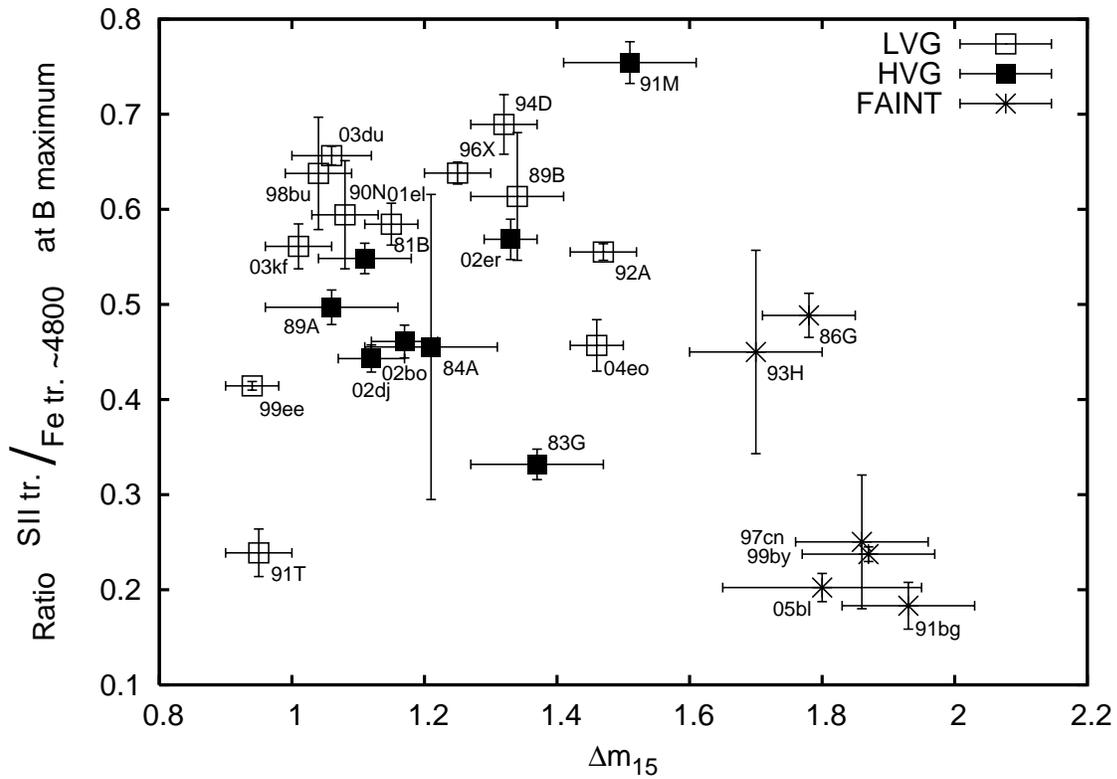}
	\caption{Ratio no. 3 -- EW(\SII\ trough)$\,/\,$EW(Fe trough $\sim4800$\,\AA) versus $\Delta m_{15}(B)$ -- IME versus Fe-dominated trough II.}
	\label{fig:R_S_Fe4800}
\end{figure}

This ratio behaves like the  previous one (Fig. \ref{fig:R_S_Fe4800}), as
could be expected since both Si and S are IME. The FAINT SNe now reach very
small values, presumably because of the higher temperature sensitivity of the
\SII\ feature than the \SiII\ $\lambda6355$ line. 

It is tempting to interpret the behaviour of this ratio and the one above as
due not only to temperature, but also to a trend for the brightest SNe to have
a higher abundance of Fe relative to IME in layers near the photosphere at
maximum  ($v \sim 10000$\,\kms). This is plausible since \FeII\ and \SiII\ have
similar ionisation potentials, and should respond similarly to changes in 
temperature. The observed behaviour may indicate that bright SNe burn more to 
nuclear statistical equilibrium (NSE) ($\sim 20$ per cent of \Nifs\ 
has decayed to \Fefs\ at the time of maximum). The
drop of the ratio at the largest $\Delta m_{15}$ values may then be due to the
fact that now the IME abundance is beginning to decrease in the mass layers
near $v_{\mathrm{ph}}$, after reaching a peak at $\Delta m_{15} \sim 1.1$--$1.5$. 

Note that $v_{\mathrm{ph}}$ is smaller at larger $\Delta m_{15}$. This implies a lower
opacity, which in turn could be associated with a smaller Fe-group abundance
relative to IME in the layers between 9000 and 11000\,\kms, that is between the
photosphere of FAINT SNe and that of the other objects. This would suggest that
the FAINT SNe produce less NSE material, as is expected both from their dimness
and their narrow light curves. The difference between FAINT SNe and brighter
ones would be in the degree of burning to NSE at velocities $\sim 10000$\,\kms,
as hypothesised in various models \citep[\eg][]{iwa99}. Burning to IME may also
extend to lower velocities in FAINT SNe than in brighter ones.

\subsection{$\mathfrak{R}$(Si,Fe) (\SiII\ $\lambda5972$ versus Fe trough $\sim4800$)}

\begin{figure}
	\centering
	\includegraphics[width=0.58\textwidth, angle=270]{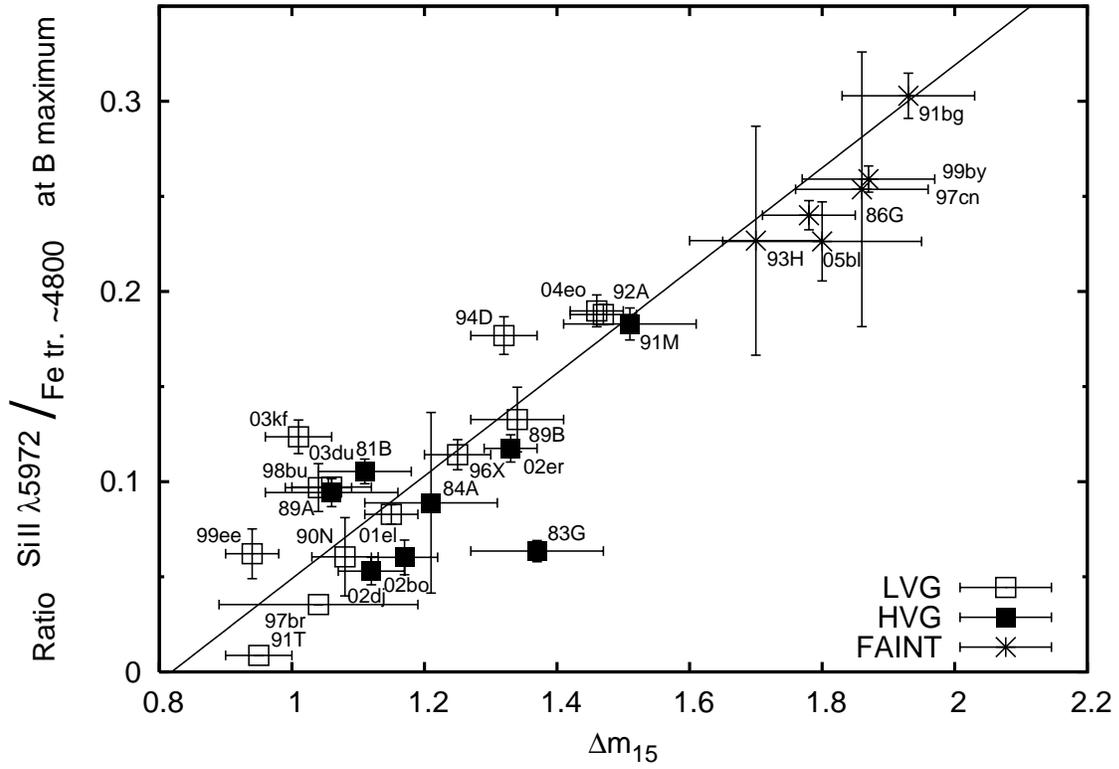}
	\caption{$\mathfrak{R}$(Si,Fe) -- EW(\SiII\ 
  $5972$\,\AA)$\,/\,$EW(Fe trough $\sim4800$\,\AA) versus $\Delta m_{15}(B)$ (91T \& 97br: upper limits): `spectroscopic L indicator'}
	\label{fig:R_Si5972_Fe4800}
\end{figure}

This ratio, unlike the previous one, shows an almost constantly rising trend.  Over a
large range of $\Delta m_{15}$ values, it increases almost linearly with $\Delta m_{15}$ 
(Fig. \ref{fig:R_Si5972_Fe4800}). This ratio is suitable as a luminosity indicator.

As for a possible explanation of the observed trend, it appears that the ratio
is driven by the increasing strength of the \SiII\ feature with increasing
$\Delta m_{15}$, which is not explained as discussed above.

\subsection{\OI\ $\lambda7773$ versus \SiII\ $\lambda6355$}

This ratio was calculated in order to investigate the relation between O and 
IME abundance. As we showed above, both EW(\SiII\ $\lambda6355$) and EW(\SII) 
decrease at  $\Delta m_{15} > 1.5$. If this implies less burning even to IME 
in the faintest SNe, we might expect O/IME ratios to increase in those objects.

\begin{figure}
	\centering
	\includegraphics[width=0.58\textwidth, angle=270]{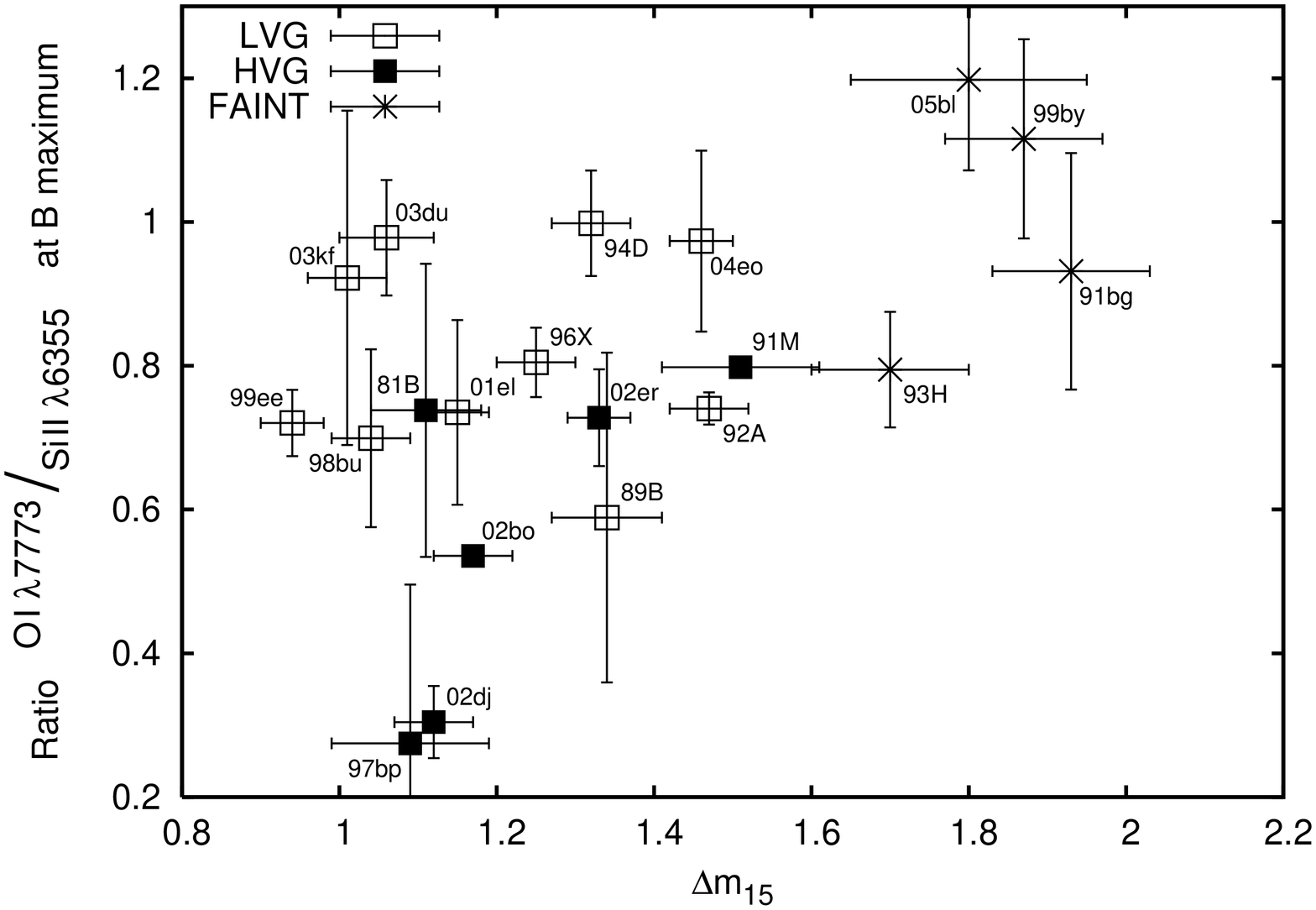}
	\caption{Ratio no. 6 -- EW(\OI\ $7773$\,\AA)$\,/\,$EW(\SiII\ $6355$\,\AA) versus $\Delta m_{15}(B)$ -- Oxygen versus IME I.\newline
	The value for 1991T (not shown) is $2.04\pm0.12$ due to its small EW(\SiII\ $6355$\,\AA).}	
	\label{fig:R_O_Si6355}
\end{figure}

The ratio of \OI\ $\lambda7773$ and \SiII\ $\lambda6355$ shows indeed a slight trend to
rise with $\Delta m_{15}$ (Fig. \ref{fig:R_O_Si6355}), but this is superimposed
by a large spread in values of $\gtrsim 25$ per cent at almost every $\Delta m_{15}$ value.
Note again that the difficulty in measuring the \OI\ line may affect our results.

\subsection{\OI\ $\lambda7773$ versus \SII\ $\lambda\sim5454,\sim5640$}

\begin{figure}
	\centering
	\includegraphics[width=0.58\textwidth, angle=270]{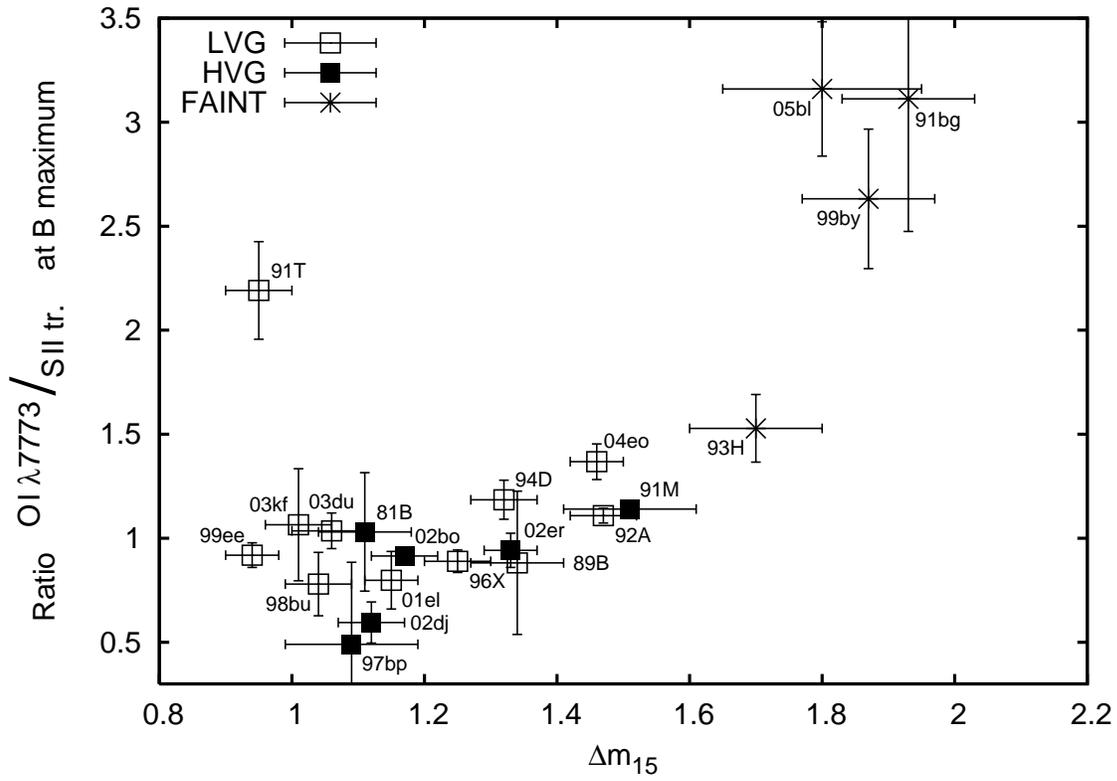}
	\caption{Ratio no. 7 -- EW(\OI\ $7773$\,\AA)$\,/\,$EW(\SII\ trough) versus $\Delta m_{15}(B)$ -- Oxygen versus IME II.}
	\label{fig:R_O_S}
\end{figure}

The \SII\ line tracks the photosphere more accurately than \SiII\
$\lambda6355$. This ratio shows tendency to increase with increasing
$\Delta m_{15}$ (Fig. \ref{fig:R_O_S}), which is enhanced for 
$\Delta m_{15}\gtrsim1.5$. While the decrease in \SII\ line strength for large 
$\Delta m_{15}$ (Fig. \ref{fig:EW_Fe4800}) certainly drives the latter trend, and the 
rise in \OI\ EW causes the tendency for $\Delta m_{15}\lesssim1.5$, how much all of 
this is due to decreasing IME abundance compared to oxygen is unclear.

\subsection{\OI\ $\lambda7773$ versus Fe trough $\sim4800$}

\begin{figure}
	\centering
	\includegraphics[width=0.58\textwidth, angle=270]{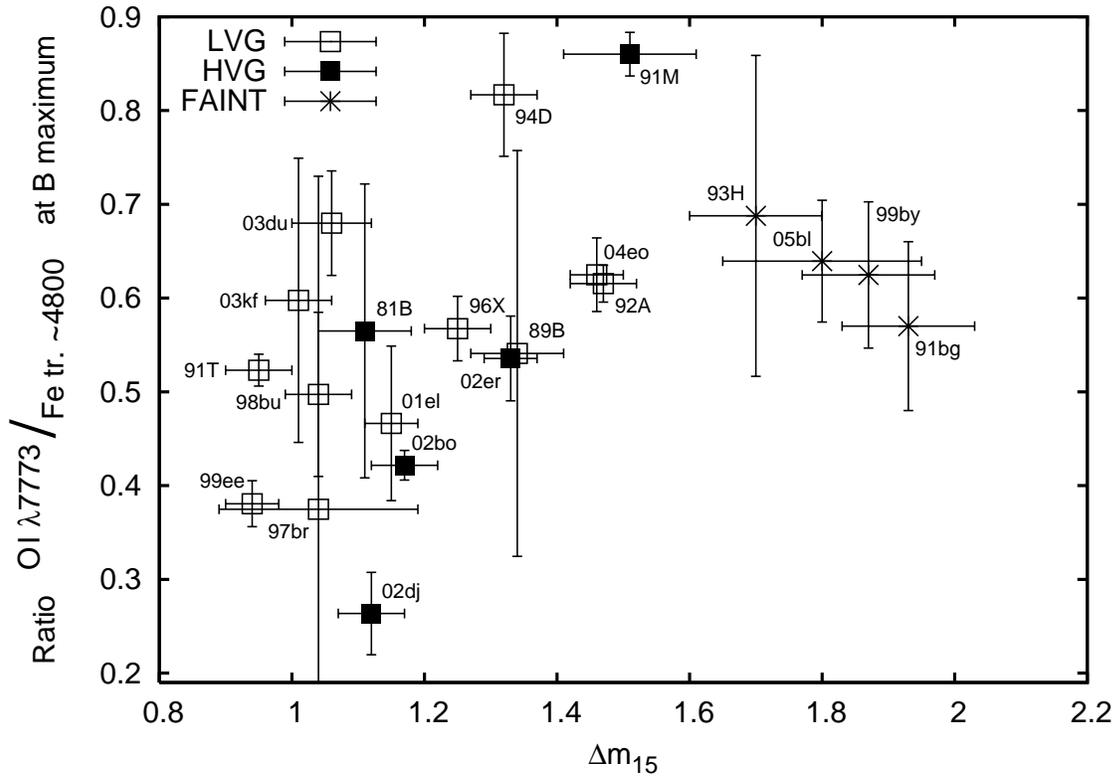}
	\caption{Ratio no. 8 -- EW(\OI\ $7773$\,\AA)$\,/\,$EW(Fe trough $\sim4800$\,\AA) versus $\Delta m_{15}(B)$ -- Oxygen feature versus Fe-dominated trough.}
	\label{fig:R_O_Fe4800}
\end{figure}

This ratio -- though exhibiting significant scatter especially at low $\Delta
m_{15}$ -- shows a clear trend to increase for $\Delta m_{15}\lesssim1.5$ (Fig. 
\ref{fig:R_O_Fe4800}). This
can be understood by  considering the tendency of the \OI\ EW to rise and the
behaviour of the Fe $\sim4800$ trough EW,  which is essentially flat.
Interestingly, for the faintest objects, an almost linear drop can be observed.

\section{Discussion}

In this section we briefly discuss the possible implications of the various
measurements.

\subsection{Photospheric velocities}

Near maximum, all LVG, some HVG and some FAINT SNe have a very similar \SiII\
velocity, $\sim 11000$\,\kms\ (Fig. \ref{fig:V_Si6355}). This can be taken to imply that there
is significant nuclear burning (at least to IME) in all these objects,
irrespective of their brightness. As we know, $\Delta m_{15}$ depends mostly on
the amount of NSE material synthesised \citep[][and references therein]{maz01},
while the kinetic energy (KE) depends also on burning to IME 
\citep{gam05}. Therefore, all LVG SNe 
may have a similar KE. The faintest SNe have a lower $v$(\SiII $\lambda6355$),
$\sim 9000$--$10000$\,\kms. This suggests that there may be less total burning, not
just less burning to NSE, and thus possibly less KE, in these SNe. 

As for HVG SNe, it is interesting to check whether the observed high velocity is
related to the presence of high-velocity features \citep[HVFs,][]{maz05b}. These
are high-velocity absorptions observed mostly in the \CaII\ IR triplet in the
spectra of almost all SNe~Ia earlier than 1 week before maximum. The high
velocities measured for HVGs here may be the result of blending of \SiII\ and
\SII\ HVFs with the lower velocity photospheric lines. Indeed, \SiII\ HVFs are
inferred at earlier times in several SNe, but never seen detached from the
main, photospheric component \citep{maz05a}.  Interestingly, no correlation
between pre-maximum HVFs and IME velocity at maximum is found: the six SNe that
are common to this study and \citet{maz05b} divide evenly among the HVG (SNe 
2002bo, 2002dj, 2002er) and LVG (SNe 2001el, 2003du, 2003kf) groups. Furthermore,
while all these SNe have prominent HVFs in the \CaII\ IR triplet about one week
before maximum or earlier, it is actually the LVG SNe among them that retain strong
\CaII\ HVFs at about maximum \citep[][Table 3]{maz05b}. 

It is reasonable to expect that detached HVFs should behave similarly, whether they
occur in \CaII\ or \SiII\ (or \SII). Therefore, the rapid decrease of the HVF
strength in HVGs may be behind the rapid drop in the \SiII\ velocity, if \SiII\
HVFs are not resolved. However, this leaves us with an apparent contradiction: on
the one hand, the LVG SNe have the longer-lasting HVFs , but on the other the 
HVG SNe still have the highest \SiII\ velocities at maximum. 
Taken individually, both of 
these behaviours could be understood in the frame of a scenario where HVFs
determine the line velocities, but the fact that they occur together is
difficult to accommodate. HVFs may be due to asymmetries in the ejection, or to
interaction with circumstellar material, while the velocity at maximum more
likely reflects global properties of the explosion. 

The \SII\ velocity behaves like the \SiII\ velocity (Fig. \ref{fig:V_S5640}). This line is
weaker than the \SiII\ line, and therefore it is a better tracer of the
photosphere. The \SII\ velocity plot shows that the photosphere moves to
progressively lower velocities for increasing $\Delta m_{15}$. This is again to
be expected, since $v_{\mathrm{ph}}$ depends on both density and opacity. While the
density may be the same, the temperature is lower in fainter SNe, so $v_{\mathrm{ph}}$
may also be lower.  The presence of S at $v \sim 7000$\,\kms\ confirms that
the \Nifs\ production is small in the faster decliners.  Small values for the
faintest SNe may also suggest a possibly smaller KE, or even a smaller mass. As
for HVGs, they may again be affected by line broadening, although clear \SII\
HVFs have never been observed. The effect is indeed smaller than seen in the
\SiII\ line, but the riddle mentioned above still stands.

\subsection{Spectroscopic luminosity indicators}

Besides $\mathfrak{R}$(\SiII), two other line strength ratios correlate
particularly well with $\Delta m_{15}$: \SII\ versus \SiII\ $\lambda5972$ 
[$\mathfrak{R}$(S,Si), Fig. \ref{fig:R_S_Si5972}] and \SiII\ $\lambda5972$ 
versus Fe $\sim4800$ [$\mathfrak{R}$(Si,Fe), Fig. 
\ref{fig:R_Si5972_Fe4800}]. All correlations involve the mysterious 
\SiII\ $\lambda5972$ line, whose EW is at least as well -- if not 
better -- correlated with $\Delta m_{15}$ than the ratios, especially 
at high values of $\Delta m_{15}$. Parameters of least 
square fits for the respective functions $\Delta m_{15}(\mathrm{ratio}|\mathrm{EW})$ can be
found in Table \ref{tab:leastsquarefittable}; the regression lines are also
shown in the respective diagrams. These linear regressions have been calculated 
over the whole SN Ia variety and not only over normal SN Ia as in \citet{bon05}.

\subsection{IME ratio differences between HVG and LVG objects}

The main difference between HVG and LVG objects, leading to the separation in a
hyerarchical cluster analysis, is the velocity development of the \SiII\
$\lambda6355$ line after maximum. The parameter $\langle\dot{v}\rangle$ seems to be 
related to the diversity of SNe Ia beyond the differences described by
$\Delta m_{15}$. HVG objects with the same $\Delta m_{15}$ exhibit a wide range
of IME velocities (Figs \ref{fig:V_Si6355} and \ref{fig:V_S5640}), 
EW(\SiII\ $\lambda6355$) (Fig. \ref{fig:EW_Si6355}), and
of the ratio EW(\SII) versus\ EW(\SiII\ $\lambda6355$) (Fig. \ref{fig:R_S_Si6355}). While the
spread of velocities could be explained by the presence of IME at different
depths in HVGs, the variation in the ratio EW(\SII)/EW(\SiII\ $\lambda6355$) is
due to the fact that only the \SiII\ $\lambda6355$ line has a wide range of EW
for the HVG SNe.

\subsection{Fe and O versus IME line strength ratios}

The line strengths around maximum give the following picture (Si conclusions
are always derived from the \SiII\ $\lambda6355$ line, as mentioned above):
brighter objects tend to contain less oxygen at the velocities probed by the
spectra near maximum (Fig. \ref{fig:EW_O}). Intermediate decliners contain more silicon
and less Fe than slow decliners (Fig. \ref{fig:R_Si6355_Fe4800}). Thus, the photosphere at maximum
is deeper in the Fe layer for the slow decliners, while it still inside the Si
layer for the intermediate decliners. However, $v_{\mathrm{ph}}$ for these two groups is
practically the same, at least within LVG objects, as shown by the $v$(Si) and
$v$(S) plots (Fig. \ref{fig:V_Si6355} and \ref{fig:V_S5640}). This implies that burning to NSE extends to
outer layers in the slow decliners.  Very faint objects contain more unburned
or partially burned material (i.e. oxygen), probably at the expense of IME
\citep[see also][]{hof02}.  This is suggested not only by the ratio EW(\OI\
$\lambda7773$)/EW(\SII) (Fig. \ref{fig:R_O_S}), but also by the decline of the equivalent
widths of the \SiII\ and \SII\ lines (Figs \ref{fig:EW_Si6355} and \ref{fig:EW_S}). Since the photosphere,
as traced by the \SII\ line, is deeper as $\Delta m_{15}$ increases, this may
suggest that the faster decliners have less overall burning.

\section{Conclusions}

We have systematically measured the velocities and EW of a number of spectral
features in SNe~Ia around maximum. The SNe have been grouped according to their
velocity gradient \citep{ben05}, and we examined different EW ratios searching for
systematic trends and for possible hints to the general character of SN~Ia
explosions. Our results can be summarised as follows. 

The photospheric velocity, as indicated by \SiII\ and \SII\ lines, is
approximately constant for all LVG SNe with $\Delta m_{15} < 1.6$. The value
declines at larger $\Delta m_{15}$. HVG SNe are found in a limited range of 
$\Delta m_{15}$, but their velocities are highly variable. 

The EW of the Fe-dominated features are approximately constant for all SNe.
Those of IME lines are highest for $\Delta m_{15} \approx 1.1$--$1.5$ and are
smaller for the brightest and the faintest SNe. HVG SNe have on average larger
values, in particular for \SiII\ $\lambda6355$. The \OI\ $\lambda7773$ line is
particularly strong in the fainter SNe, and tends to get weaker with increasing
luminosity.

Three EW ratios are good indicators of $\Delta m_{15}$: 
$\mathfrak{R}($\SiII$)$ [EW(\SiII\ $\lambda5972$)/EW(\SiII\ $\lambda6355$), 
similar to $\mathcal{R}$(\SiII) in \citet{nug95}], 
$\mathfrak{R}(\mathrm{Si,S})$ [EW(\SiII\ $\lambda5972$)/EW(\SII)], 
$\mathfrak{R}(\mathrm{Fe,Si})$ [EW(\SiII\ $\lambda5972$)/EW(Fe trough $\sim4800$)]. 
All three ratios are driven by the EW of the \SiII\ $\lambda5972$ line, which itself might 
thus be the best spectroscopic luminosity indicator. Its behaviour and
identification are, however, not well understood; these relations are therefore only
empirical. 

The ratios of EW(\SiII\ $\lambda6355$) and EW(\SII) to EW(Fe trough
$\sim4800$) (Fig. \ref{fig:R_Si6355_Fe4800} and \ref{fig:R_S_Fe4800}) show 
a parabolic behaviour: they are small at
small $\Delta m_{15}$, reach a peak at $\Delta m_{15} \approx 1.1$--$1.5$, and
then decline. While for the \SII\ line part of this behaviour could be
explained as the effect of increasing temperature, the Si/Fe trend may reflect
an abundance change. The brightest SNe have more Fe near the maximum-light
photosphere ($\sim 10000$\,\kms). Intermediate decliners have more IME and less
Fe at a similar velocity. Faint SNe have a deeper photosphere, indicating both
less \Nifs\ and Fe-group elements, and also less IME, suggesting that burning
was overall reduced. This is apprently confirmed by high \OI\ EW values
for faint SNe. 

HVG SNe have the fastest and strongest IME lines. This is, however, not correlated
with the presence of \CaII\ HVFs. Actually, SNe with the strongest, longer
lasting \CaII\ HVFs are LVGs. Longer lasting HVFs may slow down the velocity
decline, but this does not explain why among the SNe with HVFs the LVG SNe have the
lower velocities.

Our results are based on empirical measurements. It would be important to test
their implications using models. This is made complicated by the uncertainties
in the details of the abundance and density distributions, which can affect
model results. We will attempt to do this in a future work.

\section*{ACKNOWLEDGEMENTS}
This work is supported in part by the European Community's Human 
Potential Programme under contract HPRN-CT-2002-00303, `The Physics 
of Type Ia Supernovae'. We wish to thank R. Kotak, A. Pastorello, 
G. Pignata, M. Salvo and V. Stanishev from the RTN as well as 
everybody else who provided us with -- partially unpublished -- 
spectra. SH would furthermore like to thank everybody who supported 
this work at MPA. We have made use of the NASA/IPAC Extragalactic 
Database (NED, operated by the Jet Propulsion Laboratory, California 
Institute of Technology, under contract with the National Aeronautics 
and Space Administration), and the Lyon-Meudon Extragalactic Database 
(LEDA, supplied by the LEDA team at the Centre de Recherche 
Astronomique de Lyon, Observatoire de Lyon), as well as the \textsc{iraf} 
(Image Reduction and Analysis Facility) software, distributed by the 
National Optical Astronomy Observatory (operated by AURA, Inc., under 
contract with the National Science Foundation), see 
\href{http://iraf.noao.edu}{http://iraf.noao.edu}.

%%%%%%%%%%%%%%%%%%%%%%%%%%%%%%% Tables %%%%%%%%%%%%%%%%%%%%%%%%%%%%%%%%%%%%%

%%%%%%%%%%%%%%%%%%%%%%%%%%%%%%%%%%%%%%%%%%%%%%%%%%%%%%%%%%%%%%%%%%%%%%%%%%%%%%%

\begin{table*}
\begin{minipage}{133mm}
\caption{Objects and sources of respective spectra.\label{tab:objtable}}
\footnotesize
\begin{tabular}{cccl}
\hline
SN & $\Delta m_{15}(B)^{\rm a}$ & $\langle\dot{v}\rangle \left[\mathrm{km}\, \mathrm{s}^{-1}\,\mathrm{d}^{-1}\right]^{\rm a}$ & References for spectra\\
\hline
	\multicolumn{4}{c}{LVG}\\
\hline
	89B & $1.34\pm0.07$ & $66\pm5$ & \citet{bar90}; \citet{wel94} \\
	90N & $1.08\pm0.05$ & $41\pm5$ & \citet{lei91}; \citet{maz93} \\
	91T & $0.95\pm0.05$ & $11\pm5$ & \citet{fil92}; \citet{phi92}; \citet{rui92} \\
	92A & $1.47\pm0.05$ & $45\pm5^{\rm b}$ & Asiago archive; \citet{kir93} \\
	94D & $1.32\pm0.05$ & $39\pm5$ & \citet{pat96} \\
	96X & $1.25\pm0.05$ & $46\pm5$ & \citet{sal01} \\
	97br & $1.04\pm0.15$ & $25\pm5$ & Asiago archive; \citet{li99} \\
	98bu & $1.04\pm0.05$ & $10\pm5$ & Asiago archive; \citet{jha99}; \citet{her00} \\
	99ee & $0.94\pm0.04$ & $42\pm5$ & \citet{ham02} \\
	01el & $1.15\pm0.04$ & $31\pm5$ & \citet{wan03}; \citet{mat05} \\
	03du & $1.06\pm0.06^{\rm c}$ & $31\pm5$ & Stanishev et al. (2006), in preparation \\
	03kf & $1.01\pm0.05^{\rm c}$ & $50\pm5$ & Salvo et al. (2006), in preparation \\
	04eo & $1.46\pm0.04^{\rm c}$ & $45\pm4$ & Pastorello et al. (2006), in preparation \\
\hline
	\multicolumn{4}{c}{HVG}\\
\hline
	81B & $1.11\pm0.07$ & $76\pm7$ & \citet{bra83} \\
  83G & $1.37\pm0.10$ & $125\pm20$ & H83; M84; \citet{ben91}; McDonald archive\\
	84A & $1.21\pm0.10$ & $92\pm10$ & \citet{bar89} \\
	89A & $1.06\pm0.10$ & $90\pm10$ & \citet{ben91} \\
	91M & $1.51\pm0.10^{\rm d}$ & $92\pm5$ & Asiago archive; \citet{gom96}\\
	97bp & $1.09\pm0.10$ & $106\pm7$ & Asiago archive \\
	02bo & $1.17\pm0.05$ & $110\pm7$ & \citet{ben04}\\
	02dj & $1.12\pm0.05^{\rm c}$ & $86\pm6$ & Pignata et al. (2006), in preparation \\
	02er & $1.33\pm0.04$ & $92\pm5$ & \citet{kot05}\\
\hline
	\multicolumn{4}{c}{FAINT}\\
\hline
	86G & $1.78\pm0.07$ & $64\pm5$ & \citet{cri92} \\
	91bg & $1.93\pm0.10$ & $104\pm7$ & \citet{tur96}; \citet{gom96}\\
	93H & $1.70\pm0.10$ & $73\pm8$ & Asiago archive; CTIO Archive\\
	97cn & $1.86\pm0.10$ & $83\pm10$ & \citet{tur98}\\
	99by & $1.87\pm0.10$ & $110\pm10$ & \citet{gar04}\\
	05bl & $\sim1.8^{\rm e}$ & $73\pm10$ & RTN, in preparation \\
\hline
\end{tabular}
\medskip \\
\scriptsize
  $^{\rm a}$ Values from \citet{ben05} unless otherwise stated. $\Delta m_{15}(B)$ values are reddening corrected according to \citet{phi99}; \\
  $^{\rm b}$ updated value; \\
	$^{\rm c}$ private communication, preliminary values; \\
	$^{\rm d}$ see \citet{maz98}; \\
	$^{\rm e}$ estimated value from spectroscopic luminosity indicators (for regression parameters see Table \ref{tab:leastsquarefittable}) and CSP light curve (Carnegie Supernova Project, \href{http://csp1.lco.cl/~cspuser1/CSP.html}{http://csp1.lco.cl/~cspuser1/CSP.html}).\\
\normalsize
\end{minipage}	
\end{table*}

\begin{table*}
\begin{minipage}{148mm}
\caption{Overview over the features measured.\label{tab:featuretable}}
\footnotesize
\begin{tabular}{llccc}
\hline
  No.$^{\rm a}$ & Corresp. ion(s) & Rest wavelength (\AA) & Observed wavelength (\AA) & Annotations to wl. values\\
\hline
	1 & Fe--Mg(--Ti) trough$^{\rm a}$ & - & $\sim4300$ & rough estimate of centroid wl. \\
	2 & Fe trough$^{\rm a}$ & - & $\sim4800$ & rough estimate of centroid wl. \\
	3$^{\rm b}$ & \SII\ (blend) & 5454 & $\sim5250$ & rest wl.: value of strongest line \\
	3'$^{\rm b}$ & \SII\ (blend) & 5640 & $\sim5450$ & rest wl.: value of strongest (double-)line \\
	4 & \SiII\ (blend)& 5972 & $\sim5750$ & rest wl.: weighted mean \\
	5 & \SiII\ (blend)& 6355 & $\sim6100$ & rest wl.: weighted mean \\
	6 & \OI\ (blend)& 7773 & $\sim7500$ & rest wl.: weighted mean \\
\hline
\end{tabular}
\medskip \\
\scriptsize
  $^{\rm a}$ For details see text, section \ref{sec:measurements};\\
	$^{\rm b}$ features 3 \& 3': EW always measured together over the whole `W-shaped' feature
	\normalsize
\end{minipage}
\end{table*}

%%%%%%%%%%%%%%%%%%%%%%%%%%%%%%%%%%%%%%%%%%%%%%%%%%%%%%%%%%%%%%%%%%%%%%%%%%%%%%%

\begin{table*}
\begin{minipage}{108mm}
\footnotesize
\caption{Overview over the ratio values evaluated.\label{tab:ratiotable}}
\begin{tabular}{lccc}
\hline
	No. & Dividend EW & Divisor EW & Annotations\\
\hline
	1 & \SiII\ $\lambda5972$ & \SiII\ $\lambda6355$ & $\mathfrak{R}$(\SiII), similar to Nugent $\mathcal{R}$(\SiII) \\
	2 & \SiII\ $\lambda6355$ & Fe tr. $\sim4800$ & \\
	2' & \SiII\ $\lambda5972$ & Fe tr. $\sim4800$ & $\mathfrak{R}$(Si,Fe), `spectroscopic lum. indicator'\\
	3 & \SII\ tr. $\lambda 5454,5640$ & Fe tr. $\sim4800$ & \\
	4 & \SII\ tr. $\lambda 5454,5640$ & \SiII\ $\lambda6355$ & \\
	4' & \SII\ tr. $\lambda 5454,5640$ & \SiII\ $\lambda5972$ & $\mathfrak{R}$(S,Si), `spectroscopic lum. indicator'\\
	5 & Fe--Mg(--Ti) tr. $\sim4300$ & Fe tr. $\sim4800$ & \\	
	6 & \OI\ $\lambda7773$ & \SiII\ $\lambda6355$ & \\
	7 & \OI\ $\lambda7773$ & \SII\ tr. & \\
	8 & \OI\ $\lambda7773$ & Fe tr. $\sim4800$ & \\
\hline
\end{tabular}
\normalsize
\end{minipage}
\end{table*}

\begin{table*}
\begin{minipage}{80mm}
\caption{Measured Values: velocities  $\left(\mathrm{km}\:\mathrm{s}^{-1} \right)$ (at $B$ maximum). Designations in brackets refer to feature numbers in table \ref{tab:featuretable} and Fig. \ref{spectrum}. \label{tab:VelocityValueTable}}
\scriptsize
\begin{tabular}{l@{$\qquad\quad$}r@{$\quad$}r@{$\qquad\quad$}r@{$\quad$}r@{$\qquad\quad$}r@{$\quad$}r}
\hline
SN &  $v(\mathrm{F3'})$ & $\delta v(\mathrm{F3'})$ &  $v(\mathrm{F4})$ & $\delta v(\mathrm{F4})$ &  $v(\mathrm{F5})$ & $\delta v(\mathrm{F5})$ \\
\hline
\multicolumn{7}{c}{LVG}\\
\hline
89B &  9062 &   298 & 10501 &   490 & 10774 &   164 \\
90N &  9925 &   130 &  9950 &  1539 & 10598 &   128 \\
91T &  9574 &   452 & -$^{\mathrm a}$ & -$^{\mathrm a}$ & 10117 &   138 \\
92A & 10162 &    48 & 11487 &    59 & 11985 &    39 \\
94D & 10306 &    81 & 11185 &   122 & 11063 &    59 \\
96X & 10564 &    51 & 11065 &   103 & 11042 &    88 \\
97br & -$^{\mathrm a}$ & -$^{\mathrm a}$ & -$^{\mathrm a}$ & -$^{\mathrm a}$ & 11890 &   298 \\
98bu &  9855 &   250 & 10622 &   607 & 10641 &   248 \\
99ee &  9498 &   161 &  9525 &   230 & 11070 &    13 \\
01el &  9560 &   291 &  9835 &    60 & 10179 &   121 \\
03du &  9859 &    39 & 10452 &   204 & 10369 &   102 \\
03kf & 10522 &    53 & 12334 &   400 & 11349 &    20 \\
04eo &  9231 &   130 &  9939 &   189 & 10204 &   128 \\
\hline
\multicolumn{7}{c}{HVG}\\
\hline
81B & 10555 &   133 & 11321 &    43 & 11904 &    58 \\
83G & 12164 &   341 & 13095 &   211 & 15839 &   107 \\
84A & 12313 &   199 & 14810 &  1274 & 15052 &   267 \\
89A & 12453 &   381 & 13128 &   193 & 13120 &    24 \\
91M & 10314 &   161 & 11540 &   178 & 12199 &   138 \\
97bp & 12322 &   689 & 14174 &   302 & 16147 &   653 \\
02bo & 10397 &    95 & 11300 &    53 & 12942 &    58 \\
02dj & 11124 &   678 & 12342 &   272 & 13803 &   155 \\
02er & 10308 &    49 & 11332 &   196 & 11192 &   156 \\
\hline
\multicolumn{7}{c}{FAINT}\\
\hline
86G &  8091 &    88 &  9526 &    81 & 10087 &    74 \\
91bg &  7827 &   134 &  9561 &    57 & 10080 &   267 \\
93H &  8940 &   937 &  9953 &   250 & 10986 &    62 \\
97cn &  6840 &  1075 &  9456 &   278 &  9044 &   732 \\
99by &  7772 &   116 & 10052 &    19 &  9790 &    74 \\
05bl &  8706 &   312 & 10490 &    26 &  9898 &    63 \\
\end{tabular}
\medskip \\
\scriptsize
  $^{\rm a}$ Feature too weak. \\
\normalsize
\end{minipage}
\end{table*}

\begin{table*}
\begin{minipage}{160mm}
\caption{Measured Values: EWs (\AA) (at $B$ maximum). Designations in brackets refer to feature numbers in table \ref{tab:featuretable} and Fig. \ref{spectrum}. \label{tab:EWValueTable}}
\scriptsize
\begin{tabular}{l@{$\quad\;$}r@{$\quad\!\!$}r@{$\quad\;$}r@{$\quad\!\!$}r@{$\quad\;$}r@{$\quad\!\!$}r@{$\quad\;$}r@{$\quad\!\!$}r@{$\quad\;$}r@{$\quad\!\!$}r@{$\quad\;$}r@{$\quad\!\!$}r}
\hline
SN &  EW$(\mathrm{F1})\!\!\!\!$ & $\:\:\:\:\delta\,$EW$(\mathrm{F1})\!\!\!\!$ &  $\:\:\:\:$EW$(\mathrm{F2})\!\!\!\!$ & $\:\:\:\:\delta\,$EW$(\mathrm{F2})\!\!\!\!$ &  $\:\:\:\:$EW$(\mathrm{F3})\!\!\!\!$ & $\:\:\:\:\delta\,$EW$(\mathrm{F3})\!\!\!\!$ &  $\:\:\:\:$EW$(\mathrm{F4})\!\!\!\!$ & $\:\:\:\:\delta\,$EW$(\mathrm{F4})\!\!\!\!$ &  $\:\:\:\:$EW$(\mathrm{F5})\!\!\!\!$ & $\:\:\:\:\delta\,$EW$(\mathrm{F5})\!\!\!\!$ &  $\:\:\:\:$EW$(\mathrm{F6})\!\!\!\!$ & $\:\:\:\:\delta\,$EW$(\mathrm{F6})\!\!\!\!$ \\
\hline
\multicolumn{12}{c}{LVG}\\
\hline
89B &   96.7 &    5.3 &  130.7 &   13.0 &   80.2 &    3.7 &   17.3 &    1.4 &  120.1 &    4.6 &   70.7 &   27.4 \\
90N &   90.5 &    4.8 &  131.4 &   11.7 &   78.1 &    2.8 &   8.0 &   2.6 &   70.2 &    7.0 & -$^{\mathrm b}$ & -$^{\mathrm b}$ \\
91T &   88.2 &    3.2 &  129.0 &    2.0 &   30.8 &    3.2 &   1.1 &   -$^{\mathrm a}$ &   33.1 &    1.8 &   67.5 &    1.9 \\
92A &   84.0 &    1.3 &  139.8 &    1.5 &   77.6 &    0.9 &   26.3 &    0.7 &  116.2 &    0.5 &   86.0 &    2.6 \\
94D &   68.5 &    5.3 &  114.1 &    3.8 &   78.7 &    2.4 &   20.2 &    0.9 &   93.4 &    0.7 &   93.2 &    6.8 \\
96X &   93.4 &    1.4 &  130.1 &    1.7 &   83.0 &    1.0 &   14.9 &    1.0 &   91.8 &    0.8 &   73.8 &    4.4 \\
97br &   54.3 &   22.6 &  107.6 &    8.3 & -$^{\mathrm b}$ & -$^{\mathrm b}$ &   3.8 & -$^{\mathrm a}$ &   27.6 &    6.0 &   40.3 &   38.1 \\
98bu &   88.9 &    5.6 &  130.9 &    3.4 &   83.5 &    7.4 &   12.7 &    1.6 &   93.1 &    3.0 &   65.1 &   11.3 \\
99ee &  100.9 &    0.6 &  149.7 &    1.2 &   62.0 &    0.5 &   9.3 &   2.0 &   79.1 &    0.4 &   57.0 &    3.6 \\
01el & -$^{\mathrm c}$ & -$^{\mathrm c}$ &  143.8 &    5.2 &   84.0 &    1.0 &   11.9 &    0.6 &   91.2 &    2.3 &   67.1 &   11.6 \\
03du &   87.8 &    2.7 &  123.1 &    1.0 &   80.8 &    1.0 &   12.0 &    0.5 &   85.5 &    0.6 &   83.7 &    6.8 \\
03kf &   93.3 &    4.3 &  127.2 &    4.2 &   71.4 &    1.9 &   15.7 &    1.0 &   82.4 &    1.2 &  76.0 &   19.1 \\
04eo &  105.4 &   12.5 &  169.9 &    7.1 &   77.6 &    3.3 &   32.3 &    0.4 &  109.1 &   13.2 &  106.2 &    4.9 \\
\hline
\multicolumn{12}{c}{HVG}\\
\hline
81B &  110.6 &    1.2 &  168.2 &    4.3 &   92.2 &    1.3 &   17.7 &    1.0 &  128.8 &    1.3 &   95.1 &   26.2 \\
83G &  126.3 &    4.2 &  261.4 &    6.8 &   86.8 &    3.5 &   16.6 &    1.4 &  184.3 &    9.9 & -$^{\mathrm c}$ & -$^{\mathrm c}$ \\
84A &  132.3 &   18.2 &  270.0 &   12.1 &  122.9 &   43.0 &   24.0 &   12.8 &  195.1 &    1.5 & -$^{\mathrm c}$ & -$^{\mathrm c}$ \\
89A & -$^{\mathrm c}$ & -$^{\mathrm c}$ &  177.0 &    6.0 &   88.0 &    1.2 &   16.7 &    1.2 &  101.8 &    5.7 & -$^{\mathrm c}$ & -$^{\mathrm c}$ \\
91M &   92.3 &    1.2 &  121.0 &    3.0 &   91.2 &    1.4 &   22.1 &    0.9 &  130.4 &    0.7 &  104.0 &    1.1 \\
97bp & -$^{\mathrm c}$ & -$^{\mathrm c}$ & -$^{\mathrm c}$ & -$^{\mathrm c}$ &  100.3 &    8.3 &   12.3 &    1.9 &  178.7 &    7.3 &   49.1 &   39.4 \\
02bo &  104.9 &    3.3 &  184.8 &    5.8 &   85.2 &    1.7 &   11.1 &    1.6 &  145.5 &    1.7 &   77.9 &    1.6 \\
02dj &  110.5 &    2.5 &  171.8 &    3.9 &   76.1 &    1.7 &   9.1 &   1.2 &  148.8 &    1.0 &   45.3 &    7.5 \\
02er &   98.9 &    3.8 &  148.9 &    3.1 &   84.6 &    2.6 &   17.5 &    1.0 &  109.6 &    4.7 &   79.8 &    6.5 \\
\hline
\multicolumn{12}{c}{FAINT}\\
\hline
86G &  114.7 &    3.5 &  152.9 &    4.6 &   74.7 &    2.7 &   36.7 &    0.4 &  123.2 &    1.4 & -$^{\mathrm c}$ & -$^{\mathrm c}$ \\
91bg &  274.8 &   15.0 &  149.1 &    3.4 &   27.3 &    3.6 &   45.2 &    1.5 &   91.3 &    7.5 &   85.0 &   13.3 \\
93H &   97.2 &   34.4 &  139.7 &   32.0 &   62.9 &    4.0 &   31.7 &    4.3 &  120.9 &    3.2 &   96.1 &    9.4 \\
97cn &  170.7 &  116.2 &  180.6 &   35.9 &   45.2 &    8.9 &   45.8 &    9.3 &  102.0 &    9.7 & -$^{\mathrm c}$ & -$^{\mathrm c}$ \\
99by &  241.8 &    2.0 &  167.5 &    2.5 &   39.8 &    1.2 &   43.4 &    1.0 &   93.8 &    0.7 &  104.6 &   13.0 \\
05bl &  260.2 &    7.7 &  168.4 &    8.7 &   34.1 &    1.8 &   38.1 &    2.9 &   89.9 &    5.3 &  107.7 &    9.4 \\
\end{tabular}
\medskip \\
\scriptsize
  $^{\rm a}$ Values are upper limit estimates (feature very weak); $\qquad^{\rm b}$ feature too weak; $\qquad^{\rm c}$ out of spectral range.\\
\normalsize
\end{minipage}
\end{table*}

\begin{table*}
\begin{minipage}{170mm}
\caption{Measured Values: EW ratios\label{tab:RatioValueTable} (at $B$ maximum). Ratio naming refers to designations in table \ref{tab:ratiotable}.}
\scriptsize
\begin{tabular}{l@{$\qquad$}r@{$\quad$}r@{$\qquad$}r@{$\quad$}r@{$\qquad$}r@{$\quad$}r@{$\qquad$}r@{$\quad$}r@{$\qquad$}r@{$\quad$}r@{$\qquad$}r@{$\quad$}r@{$\qquad$}r@{$\quad$}r@{$\qquad$}r@{$\quad$}r@{$\qquad$}r@{$\quad$}r@{$\qquad$}r@{$\quad$}r}
\hline
SN &  $\mathfrak{R}$(\SiII) & $\delta \mathfrak{R}$(\SiII)$\!$ &  $R2\:$ & $\delta R2$ &  $\mathfrak{R}$(Si,Fe)$\!$ & $\delta \mathfrak{R}$(Si,Fe)$\!\!$ & $R3\:$ & $\delta R3$ & $R4\:$ & $\delta R4$ & $\mathfrak{R}$(S,Si)$\!$ & $\delta \mathfrak{R}$(S,Si)$\!\!$ & $R5\:$ & $\delta R5$ & $R6\:$ & $\delta R6$ & $R7\:$ & $\delta R7$ & $R8\:$ & $\delta R8$\\
\hline
\multicolumn{21}{c}{LVG}\\
\hline
89B &  0.144 &  0.013 &  0.919 &  0.098 &  0.133 &  0.017 & 0.614 & 0.067 & 0.668 & 0.040 & 4.62 & 0.43 & 0.740 & 0.084 & 0.589 & 0.229 & 0.882 & 0.344 & 0.541 & 0.216\\
90N &  0.113 &  0.039 &  0.534 &  0.071 &  0.061 &  0.021 & 0.594 & 0.057 & 1.11 & 0.12 & 9.81 & 3.25 & 0.689 & 0.071 & - & - & - & - & - & -\\
91T &  0.034 &  -$^{\rm a}$ &  0.257 &  0.015 &  0.009 & -$^{\rm a}$ & 0.239 & 0.025 & 0.930 & 0.109 & 27.35 & 13.80 & 0.684 & 0.027 & 2.04 & 0.12 & 2.19 & 0.23 & 0.523 & 0.017\\
92A &  0.226 &  0.006 &  0.831 &  0.010 &  0.188 &  0.005 & 0.555 & 0.009 & 0.668 & 0.008 & 2.95 & 0.09 & 0.601 & 0.011 & 0.741 & 0.023 & 1.11 & 0.04 & 0.615 & 0.020\\
94D &  0.216 &  0.010 &  0.818 &  0.028 &  0.177 &  0.010 & 0.689 & 0.031 & 0.842 & 0.027 & 3.90 & 0.21 & 0.600 & 0.050 & 0.998 & 0.073 & 1.18 & 0.09 & 0.817 & 0.066\\
96X &  0.162 &  0.011 &  0.705 &  0.011 &  0.114 &  0.008 & 0.638 & 0.012 & 0.905 & 0.014 & 5.59 & 0.39 & 0.718 & 0.014 & 0.805 & 0.048 & 0.889 & 0.054 & 0.568 & 0.034\\
97br &  0.138 &  -$^{\rm a}$ &  0.257 &  0.059 &  0.035 &  -$^{\rm a}$ & - & - & - & - & - & - & 0.505 & 0.213 & 1.46 & 1.41 & - & - & 0.375 & 0.355\\
98bu &  0.136 &  0.018 &  0.711 &  0.030 &  0.097 &  0.013 & 0.638 & 0.059 & 0.897 & 0.085 & 6.58 & 1.02 & 0.679 & 0.047 & 0.699 & 0.124 & 0.780 & 0.152 & 0.497 & 0.088\\
99ee &  0.117 &  0.025 &  0.529 &  0.005 &  0.062 &  0.013 & 0.414 & 0.005 & 0.784 & 0.007 & 6.68 & 1.41 & 0.674 & 0.007 & 0.720 & 0.046 & 0.919 & 0.059 & 0.381 & 0.025\\
01el &  0.130 &  0.007 &  0.635 &  0.028 &  0.083 &  0.005 & 0.584 & 0.022 & 0.921 & 0.026 & 7.06 & 0.36 & - & - & 0.735 & 0.128 & 0.798 & 0.138 & 0.466 & 0.082\\
03du &  0.140 &  0.006 &  0.695 &  0.008 &  0.097 &  0.004 & 0.657 & 0.010 & 0.944 & 0.014 & 6.75 & 0.31 & 0.714 & 0.023 & 0.978 & 0.080 & 1.04 & 0.09 & 0.680 & 0.056\\
03kf &  0.191 &  0.012 &  0.648 &  0.023 &  0.124 &  0.009 & 0.561 & 0.024 & 0.866 & 0.026 & 4.54 & 0.31 & 0.733 & 0.042 & 0.922 & 0.232 & 1.07 & 0.27 & 0.598 & 0.152\\
04eo &  0.296 &  0.036 &  0.642 &  0.082 &  0.190 &  0.008 & 0.457 & 0.027 & 0.712 & 0.091 & 2.41 & 0.11 & 0.620 & 0.078 & 0.974 & 0.126 & 1.37 & 0.09 & 0.625 & 0.039\\
\hline
\multicolumn{21}{c}{HVG}\\
\hline
81B &  0.138 &  0.008 &  0.766 &  0.021 &  0.105 &  0.007 & 0.548 & 0.016 & 0.716 & 0.012 & 5.20 & 0.30 & 0.657 & 0.018 & 0.738 & 0.204 & 1.03 & 0.28 & 0.565 & 0.157\\
83G &  0.090 &  0.009 &  0.705 &  0.042 &  0.064 &  0.006 & 0.332 & 0.016 & 0.471 & 0.032 & 5.23 & 0.49 & 0.483 & 0.020 & - & - & - & - & - & -\\
84A &  0.123 &  0.066 &  0.723 &  0.033 &  0.089 &  0.048 & 0.455 & 0.160 & 0.630 & 0.220 & 5.12 & 3.26 & 0.490 & 0.071 & - & - & - & - & - & -\\
89A &  0.164 &  0.015 &  0.575 &  0.038 &  0.094 &  0.007 & 0.497 & 0.018 & 0.864 & 0.050 & 5.27 & 0.38 & - & - & - & - & - & - & - & -\\
91M &  0.170 &  0.007 &   1.08 &   0.03 &  0.183 &  0.008 & 0.754 & 0.022 & 0.700 & 0.011 & 4.12 & 0.17 & 0.763 & 0.021 & 0.798 & 0.010 & 1.14 & 0.02 & 0.860 & 0.023\\
97bp &  0.069 &  0.011 & - & - & - & - & - & - & 0.561 & 0.052 & 8.18 & 1.47 & - & - & 0.275 & 0.221 & 0.490 & 0.395 & - & -\\
02bo &  0.076 &  0.011 &  0.787 &  0.027 &  0.060 &  0.009 & 0.461 & 0.017 & 0.586 & 0.014 & 7.66 & 1.14 & 0.568 & 0.025 & 0.536 & 0.013 & 0.915 & 0.026 & 0.422 & 0.016\\
02dj &  0.061 &  0.008 &  0.866 &  0.021 &  0.053 &  0.007 & 0.443 & 0.014 & 0.512 & 0.012 & 8.36 & 1.13 & 0.643 & 0.021 & 0.304 & 0.050 & 0.595 & 0.099 & 0.264 & 0.044\\
02er &  0.160 &  0.012 &  0.736 &  0.035 &  0.118 &  0.007 & 0.568 & 0.021 & 0.772 & 0.041 & 4.84 & 0.31 & 0.664 & 0.029 & 0.728 & 0.067 & 0.942 & 0.082 & 0.536 & 0.045\\
\hline
\multicolumn{21}{c}{FAINT}\\
\hline
86G &  0.298 &  0.005 &  0.806 &  0.026 &  0.240 &  0.008 & 0.489 & 0.023 & 0.606 & 0.023 & 2.03 & 0.08 & 0.750 & 0.032 & - & - & - & - & - & -\\
91bg &  0.495 &  0.044 &  0.612 &  0.052 &  0.303 &  0.012 & 0.183 & 0.025 & 0.299 & 0.047 & 0.605 & 0.082 & 1.84 & 0.11 & 0.931 & 0.165 & 3.11 & 0.64 & 0.570 & 0.090\\
93H &  0.262 &  0.036 &  0.866 &  0.199 &  0.227 &  0.060 & 0.450 & 0.107 & 0.520 & 0.036 & 1.99 & 0.30 & 0.696 & 0.293 & 0.795 & 0.081 & 1.53 & 0.16 & 0.688 & 0.171\\
97cn &  0.449 &  0.101 &  0.565 &  0.125 &  0.254 &  0.072 & 0.250 & 0.070 & 0.443 & 0.097 & 0.987 & 0.280 & 0.945 & 0.671 & - & - & - & - & - & -\\
99by &  0.463 &  0.011 &  0.560 &  0.009 &  0.259 &  0.007 & 0.237 & 0.008 & 0.424 & 0.013 & 0.916 & 0.033 & 1.44 & 0.02 & 1.12 & 0.14 & 2.63 & 0.34 & 0.625 & 0.078\\
05bl &  0.424 &  0.041 &  0.534 &  0.042 &  0.226 &  0.021 & 0.202 & 0.015 & 0.379 & 0.030 & 0.894 & 0.083 & 1.54 & 0.09 & 1.20 & 0.13 & 3.16 & 0.32 & 0.639 & 0.065\\
\end{tabular}
\medskip \\
\scriptsize
  \textit{Note. }Remarks regarding missing values, see table \ref{tab:EWValueTable}.\\
  $^{\rm a}$ Values are upper limit estimates (\SiII\ 5972-\AA\ feature very weak).  \\
\normalsize\end{minipage}
\end{table*}

%%%%%%%%%%%%%%%%%%%%%%%%%%%%%%%%%%%%%%%%%%%%%%%%%%%%%%%%%%%%%%%%%%%%%%%%%%%%%%%

\begin{table*}
\begin{minipage}{100mm}
\caption{Least square fit parameters and respective error values for
$\Delta m_{15}$--ratio-- resp. $\Delta m_{15}$--EW--relations 
($\Delta m_{15}=a\times \mathrm{ratio}+b$ resp. $\Delta m_{15}=a\times \mathrm{EW}+b$).\label{tab:leastsquarefittable}}
\footnotesize
\begin{tabular}{llcccc}
\hline
  Ratio & Objects excluded from fit & $a$ & $\delta a$ & $b$ & $\delta b$\\
\hline
	1 -- $\mathfrak{R}$(\SiII) & 91T$^{\rm a}$, 97br$^{\rm a}$; 05bl$^{\rm b}$ & 2.12 & 0.26 & 0.91 & 0.06\\
	2' -- $\mathfrak{R}$(Si,Fe) & 91T$^{\rm a}$, 97br$^{\rm a}$; 05bl$^{\rm b}$ ; 97bp$^{\rm c}$  & 3.71 & 0.35 & 0.82 & 0.05\\
	4' -- $\mathfrak{R}$(S,Si) & 91T$^{\rm a}$, 97br$^{\rm a}$; 05bl$^{\rm b}$& -0.104 & 0.013  & 1.83 & 0.07\\
\hline
  EW & Objects excluded from fit & $a$ & $\delta a$ & $b$ & $\delta b$\\
\hline
	F4 -- \SiII\ $\lambda5972$ & 91T$^{\rm a}$, 97br$^{\rm a}$; 05bl$^{\rm b}$ & 0.024 & 0.002 & 0.82 & 0.05\\
\hline
\end{tabular}
\medskip \\
\scriptsize
  $^{\rm a}$ \SiII\ $\lambda5972$ feature barely visible around $B$ maximum; only upper limit measurements for EW possible;\\
  $^{\rm b}$ $\Delta m_{15}$ for SN 2005bl is preliminary estimate using these correlations;\\
  $^{\rm c}$ no suitable spectrum available.\\
\normalsize
\end{minipage}
\end{table*}

\clearpage

%%%%%%%%%%%%%%%%%%%%%%%%%%%%%%%%%%%%%%%%%%%%%%%%%%%%%%%%%%%%%%%%%%%%%%%%%%%%%%%

%%%%%%%%%%%%%%%%%%%%%%%%%%%%%%% FIGURES  %%%%%%%%%%%%%%%%%%%%%%%%%%%%%%%%%%%%%

%%%%%%%%%%%%%%%%%%%%%%%%%%%%%%% Figure 1 %%%%%%%%%%%%%%%%%%%%%%%%%%%%%%%%%%%%%

%%%%%%%%%%%%%%%%%%%%%%%%%%%%%%% Figure 2 %%%%%%%%%%%%%%%%%%%%%%%%%%%%%%%%%%%%%

\clearpage
\onecolumn
%%%%%%%%%%%%%%%%%%%%%%%%%%%%%%%% Figure 3 %%%%%%%%%%%%%%%%%%%%%%%%%%%%%%%%%%%%%

%%%%%%%%%%%%%%%%%%%%%%%%%%%%%%%% Figure 4 %%%%%%%%%%%%%%%%%%%%%%%%%%%%%%%%%%%%%

\clearpage

%%%%%%%%%%%%%%%%%%%%%%%%%%%%%%%%% Figure 5 %%%%%%%%%%%%%%%%%%%%%%%%%%%%%%%%%%%%%

\clearpage
%%%%%%%%%%%%%%%%%%%%%%%%%%%%%%%%% Figure 6 %%%%%%%%%%%%%%%%%%%%%%%%%%%%%%%%%%%

%%%%%%%%%%%%%%%%%%%%%%%%%%%%%%%%% Figure 7 %%%%%%%%%%%%%%%%%%%%%%%%%%%%%%%%%%%

\clearpage

%%%%%%%%%%%%%%%%%%%%%%%%%%%%%%%%% Figure 8 %%%%%%%%%%%%%%%%%%%%%%%%%%%%%%%%%%

%%%%%%%%%%%%%%%%%%%%%%%%%%%%%%%%% Figure 9 %%%%%%%%%%%%%%%%%%%%%%%%%%%%%%%%%%

\clearpage

%%%%%%%%%%%%%%%%%%%%%%%%%%%%%%%%% Figure 10 %%%%%%%%%%%%%%%%%%%%%%%%%%%%%%%%%

%%%%%%%%%%%%%%%%%%%%%%%%%%%%%%%%% Figure 11 %%%%%%%%%%%%%%%%%%%%%%%%%%%%%%%%%%

\clearpage

%%%%%%%%%%%%%%%%%%%%%%%%%%%%%%%%% Figure 12 %%%%%%%%%%%%%%%%%%%%%%%%%%%%%%%%%%%

\clearpage

%%%%%%%%%%%%%%%%%%%%%%%%%%%%%%%%% Figure 13 %%%%%%%%%%%%%%%%%%%%%%%%%%%%%%%%%%%%

%%%%%%%%%%%%%%%%%%%%%%%%%%%%%%%%% Figure 14 %%%%%%%%%%%%%%%%%%%%%%%%%%%%%%%%%%%%

\clearpage

%%%%%%%%%%%%%%%%%%%%%%%%%%%%%%%%% Figure 15 %%%%%%%%%%%%%%%%%%%%%%%%%%%%%%%%%%%%

%%%%%%%%%%%%%%%%%%%%%%%%%%%%%%%%% Figure 16 %%%%%%%%%%%%%%%%%%%%%%%%%%%%%%%%%%%%

\clearpage

%%%%%%%%%%%%%%%%%%%%%%%%%%%%%%%%% Figure 17 %%%%%%%%%%%%%%%%%%%%%%%%%%%%%%%%%%%%

%%%%%%%%%%%%%%%%%%%%%%%%%%%%%%%%% Figure 18 %%%%%%%%%%%%%%%%%%%%%%%%%%%%%%%%%%%

\clearpage

%%%%%%%%%%%%%%%%%%%%%%%%%%%%%%%%% Figure 19 %%%%%%%%%%%%%%%%%%%%%%%%%%%%%%%%%%%%

%%%%%%%%%%%%%%%%%%%%%%%%%%%%%%%%% Figure 20 %%%%%%%%%%%%%%%%%%%%%%%%%%%%%%%%%%%

\clearpage

%%%%%%%%%%%%%%%%%%%%%%%%%%%%%%%%% Figure 21 %%%%%%%%%%%%%%%%%%%%%%%%%%%%%%%%%%%

%%%%%%%%%%%%%%%%%%%%%%%%%%%%%%%%% Figure 22 %%%%%%%%%%%%%%%%%%%%%%%%%%%%%%%%%%%

%%%%%%%%%%%%%%%%%%%%%%%%%%%%%%%%%%%%%%%%%%%%%%%%%%%%%%%%%%%%%%%%%%%

\end{document}